\documentclass[]{article}
\usepackage{epsfig}
\hoffset= - 1.0in         

\catcode`\@=11
\def\marginnote#1{}

\newcount\hour
\newcount\minute
\newtoks\amorpm
\hour=\time\divide\hour by60
\minute=\time{\multiply\hour by60 \global\advance\minute by-\hour}
\edef\standardtime{{\ifnum\hour<12 \global\amorpm={am}%
        \else\global\amorpm={pm}\advance\hour by-12 \fi
        \ifnum\hour=0 \hour=12 \fi
        \number\hour:\ifnum\minute<10 0\fi\number\minute\the\amorpm}}
\edef\militarytime{\number\hour:\ifnum\minute<10 0\fi\number\minute}

\def\draftlabel#1{{\@bsphack\if@filesw {\let\thepage\relax
   \xdef\@gtempa{\write\@auxout{\string
      \newlabel{#1}{{\@currentlabel}{\thepage}}}}}\@gtempa
   \if@nobreak \ifvmode\nobreak\fi\fi\fi\@esphack}
        \gdef\@eqnlabel{#1}}
\def\@eqnlabel{}
\def\@vacuum{}
\def\draftmarginnote#1{\marginpar{\raggedright\scriptsize\tt#1}}

\def\draft{\oddsidemargin -.5truein
        \def\@oddfoot{\sl preliminary draft \hfil
        \rm\thepage\hfil\sl\today\quad\militarytime}
        \let\@evenfoot\@oddfoot \overfullrule 3pt
        \let\label=\draftlabel
        \let\marginnote=\draftmarginnote
   \def\@eqnnum{(\theequation)\rlap{\kern\marginparsep\tt\@eqnlabel}%
\global\let\@eqnlabel\@vacuum}  }

\makeatletter
\newdimen\normalarrayskip              
\newdimen\minarrayskip                 
\normalarrayskip\baselineskip
\minarrayskip\jot
\newif\ifold             \oldtrue            \def\new{\oldfalse}
\def\arraymode{\ifold\relax\else\displaystyle\fi} 
\def\eqnumphantom{\phantom{(\theequation)}}     
\def\@arrayskip{\ifold\baselineskip\z@\lineskip\z@
     \else
     \baselineskip\minarrayskip\lineskip2\minarrayskip\fi}
\def\@arrayclassz{\ifcase \@lastchclass \@acolampacol \or
\@ampacol \or \or \or \@addamp \or
   \@acolampacol \or \@firstampfalse \@acol \fi
\edef\@preamble{\@preamble
  \ifcase \@chnum
     \hfil$\relax\arraymode\@sharp$\hfil
     \or $\relax\arraymode\@sharp$\hfil
     \or \hfil$\relax\arraymode\@sharp$\fi}}
\def\@array[#1]#2{\setbox\@arstrutbox=\hbox{\vrule
     height\arraystretch \ht\strutbox
     depth\arraystretch \dp\strutbox
     width\z@}\@mkpream{#2}\edef\@preamble{\halign
\noexpand\@halignto
\bgroup \tabskip\z@ \@arstrut \@preamble \tabskip\z@ \cr}%
\let\@startpbox\@@startpbox \let\@endpbox\@@endpbox
  \if #1t\vtop \else \if#1b\vbox \else \vcenter \fi\fi
  \bgroup \let\par\relax
  \let\@sharp##\let\protect\relax
  \@arrayskip\@preamble}
\def\eqnarray{\stepcounter{equation}%
              \let\@currentlabel=\theequation
              \global\@eqnswtrue
              \global\@eqcnt\z@
              \tabskip\@centering
              \let\\=\@eqncr
 \halign to \displaywidth\bgroup
    \eqnumphantom\@eqnsel\hskip\@centering
    $\displaystyle \tabskip\z@ {##}$%
    \global\@eqcnt\@ne \hskip 2\arraycolsep
         $\displaystyle\arraymode{##}$\hfil
    \global\@eqcnt\tw@ \hskip 2\arraycolsep
         $\displaystyle\tabskip\z@{##}$\hfil
         \tabskip\@centering
    &{##}\tabskip\z@\cr}
\begingroup\ifx\undefined\newsymbol \else\def\input#1 \endgroup\fi

\newfont{\hr}{msbm10}
\newfont{\ams}{msam10}
\font\teneufm=cmmib10
\font\seveneufm=cmmib7
\font\fiveeufm=cmmib5
\def\bfit#1{{\textfont1=\teneufm\scriptfont1=\seveneufm
\scriptscriptfont1=\fiveeufm
\mathchoice{\hbox{$\displaystyle#1$}}{\hbox{$\textstyle#1$}}
{\hbox{$\scriptstyle#1$}}{\hbox{$\scriptscriptstyle#1$}}}}

\font\numbers=cmss12
\font\upright=cmu10 scaled\magstep1
\def\stroke{\vrule height8pt width0.4pt depth-0.1pt}
\def\topfleck{\vrule height8pt width0.5pt depth-5.9pt}
\def\botfleck{\vrule height2pt width0.5pt depth0.1pt}
\def\Zmath{\vcenter{\hbox{\numbers\rlap{\rlap{Z}\kern 0.8pt\topfleck}\kern
2.2pt
                   \rlap Z\kern 6pt\botfleck\kern 1pt}}}
\def\Qmath{\vcenter{\hbox{\upright\rlap{\rlap{Q}\kern
                   3.8pt\stroke}\phantom{Q}}}}
\def\Nmath{\vcenter{\hbox{\upright\rlap{I}\kern 1.7pt N}}}
\def\Cmath{\vcenter{\hbox{\upright\rlap{\rlap{C}\kern
                   3.8pt\stroke}\phantom{C}}}}
\def\Rmath{\vcenter{\hbox{\upright\rlap{I}\kern 1.7pt R}}}
\def\Z{\ifmmode\Zmath\else$\Zmath$\fi}
\def\Q{\ifmmode\Qmath\else$\Qmath$\fi}
\def\N{\ifmmode\Nmath\else$\Nmath$\fi}
\def\C{\ifmmode\Cmath\else$\Cmath$\fi}
\def\R{\ifmmode\Rmath\else$\Rmath$\fi}

\def\e{{\,\rm e}\,}
\def\d{\partial}

\def\bea{\begin{eqnarray}}
\def\eea{\end{eqnarray}}

\def\beq{\begin{equation}}
\def\eeq{\end{equation}}
\def\ba{\beq\new\begin{array}{c}}
\def\ea{\end{array}\eeq}
\def\be{\ba}
\def\ee{\ea}
\def\stackreb#1#2{\mathrel{\mathop{#2}\limits_{#1}}}
\def\Tr{\,{\rm Tr}\,}
\def\res{{\rm res}}
\def\Bf#1{\mbox{\boldmath $#1$}}

\def\bPhi{{\Bf\Phi}}
\def\bomega{{\Bf\omega}}

\def\bsigma{{\bfit\sigma}}

\def\2{{1\over 2}}
\def\N2{${\cal N}=2$}
\def\4N{${\cal N}=4$}
\def\1N{${\cal N}=1$}

\def\eps{\epsilon}
\def\half{{\textstyle{1\over2}}}
\def\ha{{1\over 2}}

\def\pint{{-\!\!\!\!\!\!\int}}
\newcommand{\rf}[1]{(\ref{#1})}

\def\tx{{\tilde x}}
\def\LL{{\cal L}}

\newcommand{\sectiono}[1]{\section{#1}\setcounter{equation}{0}}
\renewcommand{\theequation}{\thesection.\arabic{equation}}

\begin{document}

\thispagestyle{empty}
\begin{flushright}
LPTENS-04/09\\
FIAN/TD-04/04\\
MPG/ITEP-11/04\\
IHES/P/04/05\\
UUITP-05/04\\
CTP-MIT-3472\\
\end{flushright}
\vspace{.5cm}
\setcounter{footnote}{0}
\begin{center}
{\Large{\bf Classical/quantum integrability in AdS/CFT\footnote{
\tt\noindent kazakov@physique.ens.fr\\
 \indent\ \ mars@lpi.ru, mars@itep.ru\\
 \indent\ \  joseph.minahan@teorfys.uu.se\\
\indent\ \  konstantin.zarembo@teorfys.uu.se}\par}
   }\vspace{4mm}
{\large\rm V.A.~Kazakov$^{a,}$\footnote{ Membre de l'Institut Universitaire
    de France}, A.~Marshakov$^{b,\, c,\, d}$,
J.A.~Minahan$^{e,\, f}$,
and K. Zarembo$^{e,\, c}$\\[7mm]
\large\it ${}^a$ Laboratoire de Physique Th\'eorique de l'Ecole Normale
Sup\'erieure et l'Universit\'e Paris-6,\\  
Paris, 75231, France}\\[2mm]
\large\it ${}^b$ Lebedev Physics Institute, Moscow, 119991, Russia\\[2mm]
\large\it ${}^c$ ITEP, Moscow, 117259, Russia\\[2mm]
\large\it ${}^d$ IHES, Bur-sur-Yvette, 91440, France\\[2mm]
{\large\it ${}^e$ Department of Theoretical Physics\\
Uppsala University\\
Uppsala, SE-751 08, Sweden}\\[2mm]
{\large\it ${}^f$ Center for Theoretical Physics\\
Massachusetts Institute of Technology\\
Cambridge, MA 02139, USA}\\[10mm]

{\sc Abstract}\\[2mm]
\end{center}
\noindent
We discuss the AdS/CFT duality from the perspective of integrable
systems and establish a direct relationship between the dimension of
single trace local operators composed of two types of scalar fields in
\4N super Yang-Mills and the energy of their dual
semiclassical string states in $AdS_5 \times S^5$. The anomalous
dimensions can be computed using a set of Bethe equations, which for
``long'' operators reduces to a Riemann-Hilbert problem.
We develop a
unified approach to the long wavelength Bethe equations, the classical
ferromagnet and the classical string solutions in the $SU(2)$ sector
and present a general solution, governed by complex curves endowed
with meromorphic differentials with integer periods. Using this solution
we compute
the anomalous dimensions of these long operators up to two loops and
demonstrate that they agree with string-theory predictions.

\newpage
\setcounter{page}{1}
\renewcommand{\thefootnote}{\arabic{footnote}}
\setcounter{footnote}{0}

    \sectiono{Introduction}

Gauge/string duality is an old and fascinating subject that appears in
a variety of situations \cite{Polyakov}.  This subject encompasses a
large circle of ideas, the most promising of which, perhaps, is the
relationship between strings and planar diagrams in the large-$N$
limit \cite{'tHooft:1973jz}.  A well known  example
of such a duality is the matrix model
description of  two-dimensional quantum gravity and noncritical
strings \cite{David:nj,Kazakov:ds}. Matrix models \cite{Brezin:1977sv},
together with two-dimensional QCD \cite{'tHooft:1974hx},
are two instructive examples
in which the large-$N$ limit is solvable.

A much more complicated case of this large-$N$
duality is the AdS/CFT correspondence,
an asserted equivalence of a four-dimensional gauge
theory with \4N supersymmetry and type IIB string theory
on the product space $AdS_5\times S_5$. Among other
predictions, the AdS/CFT conjecture
relates the dimensions of gauge-invariant operators
with the energies of particular closed string states propagating
in the $AdS_5\times S_5$ background
\cite{Maldacena:1998re,Gubser:1998bc,Witten:1998qj}.
At present, it has only been possible to test
this conjecture for a limited class of  operators.
The reason, of course,
is that the string calculations are trustworthy for large 't Hooft
coupling $\lambda=g^2N$, while the gauge theory calculations are
reliable if $\lambda$ is small.

There are two basic sets of operators which get around this mismatch
of the coupling strength. The first set contains
the chiral primary
operators and their descendants.
Here one may rely on nonrenormalization theorems to test directly the
correspondence \cite{Maldacena:1998re,Gubser:1998bc,Witten:1998qj}.
The second set consists of operators
with large global charges, which are dual to {\em semiclassical}
string states.
An example of this operator type are the so called BMN
operators \cite{Berenstein:2002jq}.  Here, one starts with a chiral
primary  made up of a large number of complex scalar fields
($\Tr \bPhi^L$ for $L\gg 1$), adds to the operator a handful of ``impurities'', which can
either be other scalar fields or covariant derivatives acting on the
$\bPhi$ fields, and then sums over the positions of the impurities
weighted by some phases.
On a superficial level, the dual closed string
can be visualized as a chain of scalar fields inside a trace,
 where the impurities on the chain correspond to  excitations of the
string.
The precise AdS dual of the chiral primary
is a point-like string
orbiting a geodesic of $S^5$ with angular momentum $L$,
large enough for the classical approximation to be
accurate \cite{Gubser:2002tv}.  The nearby spacetime of the string
trajectory
has a plane wave geometry in which the world-sheet string action
in the light-cone gauge can be formulated as an infinite
set of massive oscillators
\cite{Metsaev:2001bj,Metsaev:2002re}.
Therefore a direct
comparison can be made between the anomalous dimensions
of the perturbative gauge theory
\cite{Berenstein:2002jq,Gross:2002su,Santambrogio:2002sb}
and the energies of the string oscillations in the plane wave.

Since the BMN operators are assumed to have a small number of impurities,
the
strings dual to these operators are almost pointlike.  Once the string
has significant stretching,
the exact string spectrum is unknown.
Nevertheless, following  \cite{Gubser:2002tv},
a large number of classical string solutions in $AdS_5\times S^5$ have
 been constructed
 \cite{Frolov:2002av,Russo:2002sr,Minahan:2002rc,
Tseytlin:2002ny,Frolov:2003qc,Frolov:2003tu,Frolov:2003xy,
Arutyunov:2003uj,Engquist:2003rn,Arutyunov:2003za}.
These string configurations are described by solitons of the world-sheet
sigma-model, all of which have large actions and therefore correspond to
 semiclassical states that are high in the energy spectrum of the
string\footnote{
A comprehensive review of classical string solutions in $AdS_5\times S^5$
can be
found in \cite{Tseytlin:2003ii}.}. A potential difficulty in comparing
string solitons to the supersymmetric Yang-Mills (SYM) operators
is that the operators are necessarily large, {\it i.~e.}
they must contain many constituent fields
in order to carry large quantum numbers. The computation of the
anomalous
dimension for large operators, even at one loop, is hindered by operator mixing
which lifts the degeneracy of the classical scaling dimensions.
The degeneracy grows exponentially with the
size of the operators and the mixing becomes more and
more complicated. Fortunately, the mixing matrix can be identified with the
Hamiltonian
of an integrable spin chain
\cite{Minahan:2002ve,Beisert:2003jj,Beisert:2003yb}.
One can then use powerful techniques of the algebraic Bethe ansatz
\cite{Bethe:1931hc} (see the up to
date formulation of this method in \cite{Faddeev:1996iy})
to diagonalize the mixing matrix and to compute the anomalous dimensions.
Explicit calculations
have led to a
remarkable agreement with string theory predictions in several cases
\cite{Beisert:2003xu,Beisert:2003ea,Engquist:2003rn,Kristjansen:2004ei}.

Much of the recent progress focused on single trace operators
composed of two types of complex scalar fields, $\bPhi_1$ and $\bPhi_2$,
 of the form
\be\label{typop}
{\rm Tr}\left( \bPhi_1 \bPhi_1 \bPhi_1 \bPhi_2 \bPhi_2 \bPhi_1 \bPhi_2 \bPhi_2 \bPhi_1 \bPhi_1 \bPhi_1 \bPhi_2 \ldots\right)\,.
\ee
It was shown in \cite{Minahan:2002ve} that the one loop
dilatation generator when acting on operators of this  type
is
equivalent to the  Hamiltonian of the
XXX Heisenberg spin chain.  Hence, we will call operators with the form
in \rf{typop}  XXX operators.
These operators are naturally identified with the states of the
spin chain by
associating, say, $\bPhi_1 $ with up spins and $\bPhi_2 $ with down spins. For instance,
the operator \rf{typop} is mapped to the state
$$
\left|\uparrow\uparrow\uparrow\downarrow\downarrow\uparrow
\downarrow\downarrow\uparrow\uparrow\uparrow\downarrow\ldots
\right\rangle \in \left({\bf C}^2\right)^{\otimes L}.
$$
The one loop dilatation operator of \4N planar SYM theory was found
in \cite{Minahan:2002ve}, and its two loop correction was proposed in
\cite{Beisert:2003tq}.  It has the following form in this basis
\be
\label{dilatop}
\hat{D}=L+\frac{\lambda}{16\pi^2}\sum_{l=1}^L
\left(1-{\bsigma}_l\cdot{\bsigma}_{l+1}\right)+\\
\left(\frac{\lambda}{16\pi^2}\right)^2\ \sum_{l=1}^L\left(
\left(1-{\bsigma}_l\cdot{\bsigma}_{l+2}
\right)-4\left(1-{\bsigma}_l\cdot{\bsigma}_{l+1}\right)
\right) +O(\lambda^3),
\ee
where $\bsigma_l$ are the Pauli matrices which act on the spin at the
$l^{\rm th}$ site of the chain.  The dilatation operator is not
diagonal in the basis of monomials and mixes simple operators of type
\rf{typop}. It can be diagonalized, at one loop (first line
in \rf{dilatop}) by using the Bethe ansatz technique
\cite{Bethe:1931hc}. 

On the string side, restriction to  XXX operators
corresponds to a string  moving on the subspace $S^3\subset S^5$
and   localized to the center of $AdS_5$.
In \cite{Kruczenski:2003gt} it was shown how to directly compare
the equations of motion for a classical spin chain
 with the string sigma model
equations of motion by taking a ``nonrelativistic'' limit of the
effective coupling.  The classical chain naturally
corresponds to the long wave-length limit of the quantum chain
\cite{RESHSM,Smirnov:1998kv},
which is precisely what was considered in
\cite{Beisert:2003xu,Beisert:2003ea,Engquist:2003rn}.

An obvious question is how the integrability of the spin chain relates
to integrability of the sigma model.  There is a body of evidence that
the integrability of the mixing matrix for the XXX operators is not an
accidental symmetry of the one loop approximation. The XXX operators
mix only with themselves to all orders in pertubation theory and in
this subsector the integrability definitely extends to two loops
\cite{Beisert:2003tq},  to three loops
\cite{Beisert:2003tq,Beisert:2003jb,Beisert:2003ys,Klose:2003qc},
and possibly to higher loops
\cite{Serban:2004jf}.  On top of this, it has been verified that the
superstring sigma-model in $AdS_5\times S^5$ is classically integrable
\cite{Bena:2003wd,Mandal:2002fs}. Furthermore, it was shown how
solutions on the sigma model reduce to a classically integrable
mechanics, allowing one to explicitly solve for the string motion
\cite{Arutyunov:2003uj,Arutyunov:2003za}. The integrable structures on the
two sides of the AdS/CFT correspondence should be related in some
fashion.  On an algebraic level, the relation among the integrals of
motion was discussed in \cite{Dolan:2003uh}. Subsequently, Arutyunov
and Staudacher were able to make a direct comparison between
certain string solitons and large XXX operators in SYM, showing
that the {\it entire} integrable hierarchy matches in the
small coupling limit \cite{Arutyunov:2003rg}.  This matching of
hierachies has also been extended to certain classical solutions of
the $SU(3)$ chain \cite{Engquist:2004bx}.

The main goal of this  paper is to develop a unified
approach to the Bethe ansatz, string solitons and the weak-coupling
``non-relativistic'' reduction of the sigma-model. We will discuss in detail
mainly the XXX subsector, though we will also mention a set of
string solutions that are not dual to
this class of operators.

In comparing perturbative gauge theory results to string theory
predictions, it is important to note that gauge theory and
string theory compute different expansions  for the anomalous
dimension.
The
string calculations are accurate when both the bare dimension of the
operator, $L$, and the 't~Hooft coupling $\lambda$ are large, but the
ratio $\lambda/L^2$ is finite.  Since the relevant parameter is
$\lambda/L^2$, it is possible to consider strong coupling on the
string side and still have $\lambda/L^2$ small.  One might then expect
to make a direct comparison to perturbative gauge theory.  

However,
it was argued by Serban and Staudacher \cite{Serban:2004jf} that
nonperturbative terms could complicate the situation. These authors
showed that a particular long range spin chain would reproduce the
two-loop and three-loop dilatation operators of \cite{Beisert:2003tq,Beisert:2003jb,Beisert:2003ys} if one identifies $\lambda/L^2$ with a parameter
of the spin chain.  They then went on to show that the operators dual
to a folded and a circular
string match at two-loops, but fail to do so at three-loops.

But even  two-loop matching is mysterious.
On general grounds, the classical string
predicts the dimension of the operator to be
$\sqrt{\lambda}F(\lambda/L^2)$
with some function $F(\lambda/L^2)=L/\sqrt{\lambda}+...$.
Any stringy nonperturbative contribution to the
dimension should be of the form
\be\label{nonpert}
f(\lambda/L^2)\exp(-\sqrt{\lambda}g(\lambda/L^2)).
\ee
Of course this term and others like it cannot be in general separated in
the same exponential form on the gauge side, but just as an illustration
let us momentarily take them seriously, in which case they can be perturbative
in the gauge theory.
In order to
have a sensible contribution to perturbative gauge theory, we should
require that the Taylor expansion of $f(\lambda/L^2)$ and
$g(\lambda/L^2)$ about $\lambda/L^2=0$ have integer and half integer
powers respectively.
If the lowest power of $g$ is $n\leq -\half$, then there would be an
exponential suppression in $L$ for all orders of $\lambda$ and
(\ref{nonpert}) would not contribute to any order in perturbation
theory in the large $L$ limit. However, if the first power is $n=\half$,
then
(\ref{nonpert}) {\em can affect} the perturbative expansion in
the large $L$
limit.  However, (\ref{nonpert}) should not contribute to the bare
dimension of the operator, requiring that $f(0)=0$.  Hence, the term
that can affect the perturbative gauge theory expansion at the lowest
possible order would have the form $\sim
(\lambda/L^2)(1-a\lambda/L+\ha a^2\lambda^2/2L^2+...)$.  The lowest
term in this expansion is a one-loop term, but it is suppressed by a
factor of $L$ from the classical string contribution so it can be
ignored.  The second term is a two loop term which scales like the
classical contribution and so its existence would affect the two loop
gauge contribution.  The third term is a three-loop term which does
not even have perturbative BMN scaling.  What is puzzling is
that the breakdown described here seems
to take place at one loop higher.  In fact it was shown in
\cite{Serban:2004jf} that assuming that perturbative Yang-Mills is
described by an Inozemtsev chain \cite{Inozemtsev},
perturbative BMN scaling breaks down at four loops.

We will demonstrate that indeed the Riemann-Hilbert problem on the
string theory
side immediately reproduces  one and two loops in gauge theory.
We do not study the three-loop reduction of the sigma-model
in this paper, but are planning ot return to this question in
the near future.

In what follows we will consider generic solutions to the class of
so called finite-gap potentials.  These solutions are
directly related to
 algebraic geometry on  finite genus
Riemann surfaces. From the point of view of the Bethe equations,
this corresponds to the Bethe roots condensing onto a finite
number of ``supports" in the complex plane. Such geometric objects have
 been previously used in SUSY gauge theories.
For example, the Seiberg-Witten solution to
\N2 SYM is formulated in terms of the geometry of curves \cite{SW}
which in turn can be
nicely translated into the language of integrable systems \cite{GKMMM}
in a class closely related to those considered below.
In the \N2 case
 the trajectories of the integrable model
correspond  to BPS states while in the \4N case
they correspond to composite operators.  This relationship
deserves
further investigation.

We should mention that integrable spin chains arise
in large-$N$ QCD
\cite{Lipatov:1993yb,Faddeev:1994zg,Braun:1998id,Braun:1999te,Belitsky:2003ys}
(the quasiclassical solutions to the Bethe anzatz equations were
considered in this context in \cite{KorchQ,KorchKri})
and in less
supersymmetric cousins of \4N SYM
\cite{Wang:2003cu,Stefanski:2003qr,Roiban:2003dw,Chen:2004mu,DeWolfe:2004zt}.
Potential relevance of the XXZ spin chain
for gauge/string theory duality was discussed in
\cite{Gorsky:2003nq}.

In sect.~\ref{ss:bethe} we review the Bethe ansatz and discuss the
reduction of the Bethe equations in the scaling limit to integral
equations, similar to those found for matrix
models.  In sect.~\ref{ss:generic} we construct the generic solution
of this Bethe ansatz integral equation describing the one-loop \4N
Yang-Mills theory and then generalize it to two loops.
Sect.~\ref{ss:classic} is devoted to classical integrability: we show
that the solutions of the Bethe equations in the scaling limit are in
one-to-one correspondence with the finite-gap solutions of the
weak-coupling sigma-model at one loop in the gauge coupling. Moreover, we
then derive the classical counterpart of the Bethe equations in the
string sigma-model and discuss the general underlying geometry of the
sigma-model solutions. In particular, we find that the resolvent of
the Bethe roots, which is the generator of the higher charges of the
integrable system, is closely related to the quasi-momentum of the
sigma-model and that the latter reduces to the former in the limit of
zero coupling, {\it i.e.} at one loop.  At two loops, the precise map between
the sigma-model and the spin chain involves redefinition of the spectral
parameter.  In
section \ref{ss:examples} we apply the general formalism of
sections \ref{ss:generic} and \ref{ss:classic} to concrete
examples. We obtain some new solutions as well as rederive known
results to illustrate the general method. Included in these examples
are general rational solutions where the Bethe roots condense onto one
cut.
We also consider a
sigma-model solution
which corresponds to a ``pulsating string'' and show that the
quasi-momentum reduces to the resolvent of a particular solution of
the $SO(6)$ chain in the scaling limit.  In sect.~\ref{ss:discussion}
we present conclusions and speculations. Some technical issues are presented
in Appendices.

\sectiono{The Bethe Ansatz}
\label{ss:bethe}

Here we will derive the Bethe-type equations for the diagonalisation
of the one-loop dilaton operator. The generalisation to two loops will
be presented at the end of the section 3.

If we assign $R$-charges $(1,0)$ and $(0,1)$ to the complex scalars $\bPhi_1 $
and $\bPhi_2 $, the eigenstates of the Heisenberg Hamiltonian with $J$ down spins
out of $L$ total spins
correspond to conformal operators with $R$-charges $(J_1,J_2)=(L-J,J)$:
\be
{\cal O}={\rm Tr}\left(\bPhi_1 ^{L-J}\bPhi_2 ^J+{\rm permutations}\right).
\ee
Each eigenstate of the dilatation operator \rf{dilatop}
describes a collection of $J$ interacting spin waves
with rapidities $u_1,\ldots, u_J$. The spin waves attract and can form bound
states, so the rapidities are in general complex. The eigenstates
form representations of $SU(2)$ generated by the total
spin which commutes with the Hamiltonian. The highest weight in each $SU(2)$
representation can be
constructed with the help of the algebraic Bethe ansatz \cite{Faddeev:1996iy}
and is
completely characterized by a set of rapidities.  These rapidities
must be distinct
complex numbers, are either real or form  complex conjugate pairs,
and satisfy a set of  algebraic Bethe equations.

The integrable structure of the Heisenberg spin chain is encoded in the
transfer
matrix (see, for example, \cite{Faddeev:1996iy} and references therein)
defined as
\be
\label{trama}
\hat{T}(u)={\rm Tr}\,\left[\left(u
+\frac{i}{2}\,{\bsigma}_L\otimes{\bsigma}\right)
\ldots
\left(u
+\frac{i}{2}\,{\bsigma}_1\otimes{\bsigma}\right)\right].
\ee
The Pauli matrices without a superscript $\bsigma$ act in a two-dimensional
auxiliary space, and tracing in (\ref{trama}) over the auxiliary
space leaves an operator in the Hilbert space
of the spin chain. The transfer matrices at
different values of the spectral parameter $u$ commute and therefore
generate an infinite set of conserved charges in involution
\footnote{To be more precise, for the chain of length $L$
there are exactly $L$ independent charges,
since the transfer matrix is a polynomial of
order $L$.}.
The Taylor expansion of the transfer matrix at $u=i/2$
generates mutually commuting
local charges:
\be\label{tu}
\hat{T}(u)=\left(u+\frac{i}{2}\right)^L

\hat{U} \exp\left[i\sum_{n=1}^{\infty}
\frac{1}{n}\,\left(u-\frac{i}{2}\right)^n\hat{Q}_n\right],
\ee
where $\hat{U}=\e^{i\hat{P}}$ is the shift operator that
generates translations by one site.
The first charge from the sum in the exponent of (\ref{tu})
is proportional to the one-loop dilatation operator \rf{dilatop}:
$\hat{D}=L+{\lambda\over 8\pi^2}\,\hat{Q}_1+O(\lambda^2)$.
The Laurent expansion of (\ref{trama}) at $u=\infty$
(before taking the trace over the auxiliary matrices)
produces the total spin and non-local Yangian charges (see the
recent discussion of Yangian symmetry and its relation to Yang-Mills
theory in \cite{Dolan:2003uh}).

The Bethe ansatz diagonalizes the whole tranfer matrix,
\be
\hat{T}(u)\left|u_1\ldots u_J\right\rangle
=T(u)\left|u_1\ldots u_J\right\rangle,
\ee
giving eigenvectors that depend on $J$ rapidities and whose
eigenvalues are
\be\label{eigenv}
T(u)=\left(u+\frac{i}{2}\right)^L\prod_{k=1}^J\frac{u-u_k-i}{u-u_k}
+\left(u-\frac{i}{2}\right)^L
\prod_{k=1}^J\frac{u-u_k+i}{u-u_k}
\ee
The second term does not affect the local charges \rf{tu} as it
is multiplied by a huge power of $u-i/2$.

The Bethe equations are obtained from \rf{eigenv} by
simply noting that the function $T(u)$, by
construction, must be a polynomial of degree $L$, or for
$L\to\infty$ an entire analytic
function of $u$. Hence it has no poles at any $u=u_j$, and this
immediately leads to the $J$ conditions
\be\label{BAEQ}
\left(u_j+i/2\over u_j-i/2\right)^L=
\prod_{k=1(k\ne j)}^J {u_j-u_k+i\over u_j-u_k-i},
\ee
which are in fact the Bethe equations.

 These Bethe equations also represent periodicity conditions for the
 wave function and can be written in the logarithmic form as
\be\label{BAE}
L\log\left(u_j+i/2\over u_j-i/2\right)=
\sum_{k=1(k\ne j)}^J \log{u_j-u_k+i\over u_j-u_k-i}
-2\pi in_j.
\ee
The mode numbers $n_j$ are arbitrary integers.
The cyclicity of the trace in
the operators of \4N supersymmetric Yang-Mills requires that
\be
\label{MOMu}
\e^{iP}=\prod_{j=1}^J{u_j+i/2\over u_j-i/2}=1,
\ee
which is the quantization condition on the total momentum: ${p\over
2\pi}\in\Z$.  The energy of the solution gives the anomalous dimension
$\gamma$ of the corresponding operator in
Yang-Mills theory \cite{Minahan:2002ve},
\be
\label{gamad}
\gamma={\lambda\over 8\pi^2}\sum_{j=1}^J{1\over u_j^2+1/4}\,.
\ee
In the scaling
limit of large operators where $L\to\infty$, all Bethe roots are of  order
$u_j\sim L$.  By rescaling $u_j = Lx_j$ we can then write
eq. (\ref{BAE}) in the form
\be\label{BAEG}
{1\over x_j}={2\over L}\sum_{k=1(k\ne j)}^J {1\over
x_j-x_k}-2\pi n_j.
\ee

Let us consider the situation with a
finite number of different
mode numbers $n_j$, and
assume that the number of Bethe roots with the same mode
number is of order $L$. The roots with mode number $n_k$ are
centered roughly around the point $x= 1/(2\pi n_k)$, with a typical
separation between adjacent roots of order $\Delta x\sim 1/L$.   Hence,
roots with the same mode number form a continuous contour in the
complex plane in the scaling limit $L\rightarrow\infty$\footnote{The
limit we consider here is different from the standard thermodynamic
limit. We are keeping mode numbers $n_i$ finite, while in the
thermodynamic limit, momenta $p_i= 2\pi n_i/L$ are fixed as $L$ goes
to infinity. Our Bethe equations are reminiscent of those for the so
called Gaudin model.}. 

Equations (\ref{BAEG}) are very similar to the
saddle point equations for matrix models (apart from the last term, to
be discussed below),
where the Bethe roots are the analogs of the matrix eigenvalues.  Just
like for matrix eigenvalues, one can characterize the distribution of
Bethe roots by the density
\be
\rho(x)=\frac{1}{L}\sum_j\delta(x-x_j)
\ee
or by the resolvent
\be
\label{defG}
G(x ) = {1\over L}\sum_{j=1}^J {1\over x-x_j} =
\int_{\bf C}{d\xi\,\rho(\xi)\over x-\xi}\,.
\ee
The density is non-zero on a set of contours ${\bf C}$ in the complex plane
which in general consists of several
non-overlapping cuts,
${\bf C}={\bf C}_1\cup\ldots\cup{\bf C}_K$,
where the $l^{\rm th}$ cut ${\bf C}_l=(a_l,b_l)$ represents roots with  mode
number $n_l$.
Though we will assume that the number of cuts is finite, this
requirement can certainly be relaxed. We will return to this point in
the discussion of classical integrability of the string sigma-model.
As follows from its definition,  the resolvent
 is an analytic function on  
the complex plane with cuts. Asymptotically it has the form
\be
\label{asinf}
G(x)=\frac{\alpha}{x}+\dots\,,~~~~~(x\rightarrow\infty),
\ee
which reflects the normalization condition for the density
\be
\label{RHON}
\int_{\bf C}d\xi\,\rho(\xi)=\alpha,
\ee
where $\alpha=J/L$ is the filling fraction, the fraction
of down spins in the Bethe state.
Strictly speaking, Bethe states only make sense for $\alpha\leq \half$, but
it was observed in a number of examples
\cite{Beisert:2003xu,Beisert:2003ea,Engquist:2003rn}
that solutions of the Bethe equations can be analytically continued past
$\alpha=\half$.
We will clarify the meaning of this analytic continuation in
sections ~\ref{ss:generic}
and \ref{ss:ellip}. The analytic continuation is always possible and
 the analytically continued solution
corresponds to some Bethe state with $\alpha$
replaced by $1-\alpha$. Since replacing $\alpha$ with $1-\alpha$
is equivalent to interchanging $\bPhi_1 $ and $\bPhi_2 $ in the operator
${\rm Tr}(\bPhi_1 ^{J_1}\bPhi_2 ^{J_2}+\ldots)$, $J_1=L-J$, $J_2=J$, the resulting state has
essentially the same charges. We thus allow for
$0\leq \alpha\leq 1$ and regard charges of the states
with $\alpha$ and $1-\alpha$ as equal.

The scaling limit of the Bethe equations (\ref{BAEG}) can be
rewritten as an integral equation for the density,
\be\label{BAEC}
G(x+i0)+G(x-i0) =
2 
\pint_{{\bf C}} {d\xi\,\rho(\xi)\over x-\xi}
= {1\over x}+2\pi n_l,\ \ \ x\in
{\bf C}_l
\ee
The momentum condition (\ref{MOMu}) after rescaling  turns into
\be
\label{MOMx}
P={1\over L}\sum_{j=1}^J {1\over x_j} = 2\pi m, \ \ \ \ m\in \Z
\ee
or, 
taking into account the Bethe equations (\ref{BAEG}) and  (\ref{BAEC}),
\be
\label{MOMrho}
-{1\over 2\pi i}\oint_{{\bf C}}{G(x)dx\over x} =
\sum_{k=1}^K \int_{{\bf C}_k}{\rho(x)dx\over x}
= 2\pi \sum_{l=1}^K n_l \int_{{\bf C}_l}\rho(x)dx
= 2\pi m, \ \ \ \ \ n_l, m\in \Z.
\ee
For the anomalous dimension (\ref{gamad}) we obtain in the
scaling limit
\be
\label{gamacl}
\gamma={\lambda\over 8\pi^2L}\int dx\,\frac{\rho(x)}{x^2}
=-{\lambda\over 8\pi^2 L}\oint_{\bf C} {dx\over 2\pi i\, x^2}\, G(x)
\ee
Rescaling the spectral parameter $u=Lx+i/2$ in the eigenvalue of the
transfer matrix and dropping the second irrelevant term in
\rf{eigenv}
we find in the limit $L\rightarrow\infty$:
\be\label{retrama}
\frac{T(Lx+i/2)}{\left(Lx+i/2\right)^L}
=\exp\left(-\frac{i}{L}\sum_j\frac{1}{x-x_j}\right)
\ee
Comparing with (\ref{tu}), we see that the resolvent (\ref{defG})
is the generating function of local
charges in the scaling limit \cite{Arutyunov:2003rg,Engquist:2003rn}.
Obviously, the first two coefficients in its Taylor expansion
are the total momentum and the energy.
\footnote{ There are only $L$ independent charges, so only the
first $L$
coefficients of the $T(u)$ Taylor expansion make sense as conserved
charges. Unless we are interested in non-local charges, we can forget
about the second term in \rf{eigenv}.
The generating function of the local
charges is a pure phase: $\exp\left(i {\rm Hermitean\ operator}\right)$
and not a
cosine of it. In sect.~\ref{ss:classic} we get 
a cosine of a Hermitean operator
because $T(u)$ is
expanded at the ``wrong'' point $u=0$ (instead of $u=i/2$). In effect,
the argument of the cosine is shifted by $\half x$ compared to the argument of
the exponential.}

Our derivation of the integral equation for the density of Bethe roots
rested on the assumption that the right-hand side of the Bethe
equations can be expanded in powers of $1/(u_j-u_k)$.  This is not
always true. The assumption breaks down if the distance between nearby
roots $u_j- u_k$ approaches $i$. Some solutions of the Bethe equations
are known to have roots separated exactly by $i$ in the thermodynamic
limit. The roots form arrays with equidistant separation: $u_j= u_0
+i\ j $. Such arrays are usually referred to as ``strings''
\cite{Faddeev:1996iy} and survive the thermodynamic limit due to the
chain cancellation of dangerous terms in the sum entering
eq. (\ref{BAE}). It is certainly possible to have large strings
composed of a macroscopic number of roots.  The density will then have
constant pieces with a flat distribution, $\rho(x)=1$. Such
condensates of Bethe roots necessarily have tails with smaller
density. Condensates thus can be viewed as extra logarithmic cuts in
the complex plane that connect pairs of the original contours ${\bf
C}_l$ (the situation is identical to that arising in multicolor 2D
Yang-Mills theory on the sphere \cite{Douglas:1993ii}). The resolvent
experiences a jump by $2\pi $ across any of these extra cuts and is no
longer a single-valued function.  The integral equation (\ref{BAEC})
stays the same, but we have extra conditions on its solution as it
crosses these logarithmic cuts.

The solution of the Bethe equations with a condensate along the
imaginary axis is one of the examples in which the agreement between
the anomalous dimension and the energy of a string soliton was first
observed \cite{Beisert:2003xu}.  In that example, however, the roots
were separated not by $i$, but by a half-integer interval $i/2$, and
the condensate had the density $\rho(x)=2$.  We do not see any reason
why strings of an arbitrary integral density cannot appear: for the
distribution of roots $u_j= u_0 +i\ {j\over m}$, where $m$ is any
positive integer, the same cancellation mechanism is at work
\cite{MATTHIAS}. Though these denser strings apparently have never
been discussed in the literature on the Bethe Ansatz, their existence
is quite natural, at least in the long chain limit considered
here. Another argument in favor of condensates of an arbitrary density
is that solutions $\rho(x)=2m$ are necessary to match the string
theory predictions, where the integer $m$ plays the role of the
winding number.

To summarize, Bethe roots in the scaling limit are continuously
distributed on a collection of contours ${\bf C}_l$ in the complex
plane and may also form condensates, which connect pairs of contours
and on which the density takes constant integer values.  The density
of roots $\rho(x)dx$ is real, positive and is normalized to the
filling fraction $\alpha$. It must also be symmetric under complex
conjugation of $x$.  The density of Bethe roots satisfies an integral
equation (\ref{BAEC}) which holds on each of the ${\bf C}_l$'s.  All
possible solutions to the Bethe equations in the scaling limit are
thus parameterized by specifying mode numbers $n_l$ associated with
contours ${\bf C}_l$ and integer densitities of condensates
$m_l$. Once these numbers are specified we can solve the integral
equation, at least in principle, and find the density.  We will see,
however, that there is a symmetry that transforms different sets of
integers $(n_l,m_l)$ into each other so that solutions with different
sets of $m$'s and $n$'s lead to the same resolvent.

\sectiono{General solution}
\label{ss:generic}

In this section we first construct the general solution of
equation (\ref{BAEC}) and then consider corrections to this solution
arising from the two-loop contributions in SYM theory.

\subsection{ General one-loop solution }

Following the procedure proposed in \cite{RESHSM} we introduce
\be\label{pofG}
p(x)=G(x)-{1\over 2x},
\ee
which we shall call the quasi-momentum for reasons explained in the next
section, and rewrite (\ref{BAEC}) in the form of a Riemann-Hilbert
problem defining the analytic function $p(x)$ from its
discontinuities
\be
\label{eqxxx}
p(x+i0)+p(x-i0) =  2\pi n_i,
\ \ \ \ \ \ \ \ x\in {\bf C}_i
\ee
on every cut ${\bf C}_i$. Equation \rf{eqxxx} together
with relation \rf{pofG} are very similar
to the corresponding equations arising in the quasiclassical
computation of matrix integrals. The main difference is the
presence of the mode numbers $n_j$, which are different on the different
eigenvalue supports. Therefore, unless there is only one cut,
the resolvent $G(x)$ and the quasimomentum
$p(x)$ are not algebraic functions, i.e. in contrast to a very similar
equation found for the one matrix model (see e.g. \cite{David:sk}) and even
the
more complicated situation of the two-matrix model \cite{KM},
the resolvent
and quasimomentum can {\em not} be determined
by solving a polynomial equation~\footnote{This is
a crucial difference between
the quasiclassical solutions to the Bethe equations we consider below,
as compared
to those of \cite{KorchQ}, which correspond to the limit $J\to\infty$ at
{\em finite} $L$. In the last case the resolvent and quasimomentum are
logarithms of algebraic functions and there is no freedom in
choosing discontinuties in \rf{eqxxx}, while below, together with the
filling fractions of eigenvalues, these will be the
parameters of our general solution.}.

It follows nevertheless from (\ref{eqxxx}) that $p(x)$ is a double-valued
function of $x$. It must be single-valued
on the physical sheet of its Riemann surface, which is a complex
plane with cuts ${\bf C}_j$. The only singularities of $p(x)$
are two poles at $x=0$ on the two sheets of the Riemann
surface. The pole on the physical sheet should cancel in $G(x)$,
which by definition should be regular at zero. In other
words, $p(x)=\int^x dp$ is an Abelian integral for the meromorphic
differential $dp$, which has two double-poles at $x=0$ and
integer periods on a hyperelliptic curve $\Sigma$. Let
\be
\label{sigmaxxx}
\Sigma: \ \ \ \ y^2 = R_{2K}(x) = x^{2K} +
r_1x^{2K-1} + \dots + r_{2K} = \prod_{j=1}^{2K} (x-{\rm x}_j)
\ee
where the product is taken over the end-points of all $K$ cuts ${\bf
C}_l\in {\bf C}$.  We should also impose the reality constraint on the
whole solution such that
the branch-points of every cut have complex
conjugated coordinates, ${\rm x}_{2j}={\rm x}_{2j+1}^*$.
The other possiblity is
that some ${\rm x}_k$'s are real, but it is never realized in the
present case, although solutions with real cuts do exist
for spin $s=-\half$ chains \cite{Beisert:2003ea} and are relevant for
SYM duals of
strings with nonzero angular momentum in  $AdS_5$ \cite{Frolov:2002av}.
In another context the quasiclassical solutions to Bethe equations with the
cuts along the real axis were discussed in \cite{KorchQ}.

\begin{figure}[tp]
\centerline{\epsfig{file= 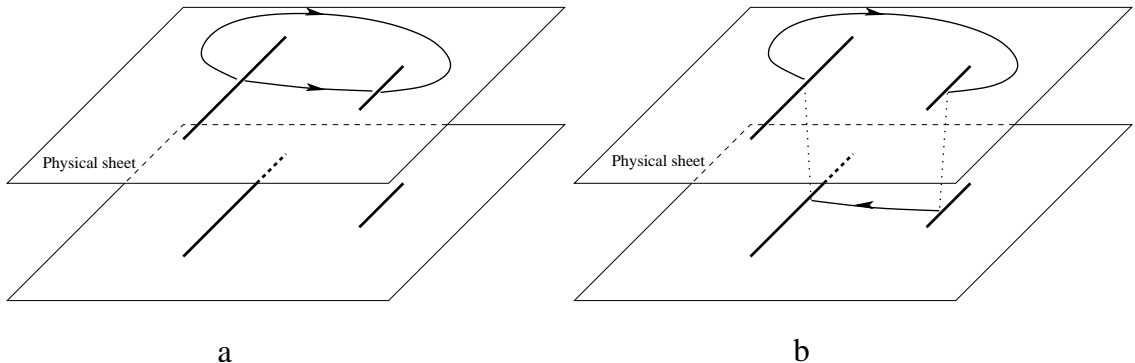,width=150mm}}
\caption{Contours of integration in eq.~(\ref{der}):
on the physical sheet (a) and after we move the contour into
the second sheet with simultaneous flip of orientation (b).}
\label{flipc}
\end{figure}

Any meromorphic differential on \rf{sigmaxxx} has general structure
$df(x) + {g(x)dx\over y}$,
where $f(x)$ and $g(x)$ are two rational functions. From
(\ref{eqxxx}) one can expect for that differential $dp$ is just
$dp=g(x)\,\frac{dx}{y}$, and
we can rewrite (\ref{eqxxx}) in terms of integrals over the
${\bf B}$-cycles on $\Sigma$,
\be
\label{der}
2\pi (n_i-n_j) =
p(x_i+i0)-p(x_j-i0)+p(x_i-i0)-p(x_j+i0) =
\\
= \int_{x_j-i0}^{x_i+i0}dp +
\int_{x_j+i0}^{x_i-i0}dp
= \oint_{B_{ij}}dp.

\ee
Here we used the hyperelliptic involution $y\rightarrow -y$ of the
Riemann surface
$\Sigma$ to move the contour of integration from the physical sheet
to the second sheet. The involution interchanges the two
sheets and flips the sign of $y$, then the differential
$dp$ changes sign too  and hence we should
simultaneously reverse the
orientation of the contour, this is illustrated in fig.~\ref{flipc}.

\begin{figure}[tp]
\centerline{\epsfig{file= 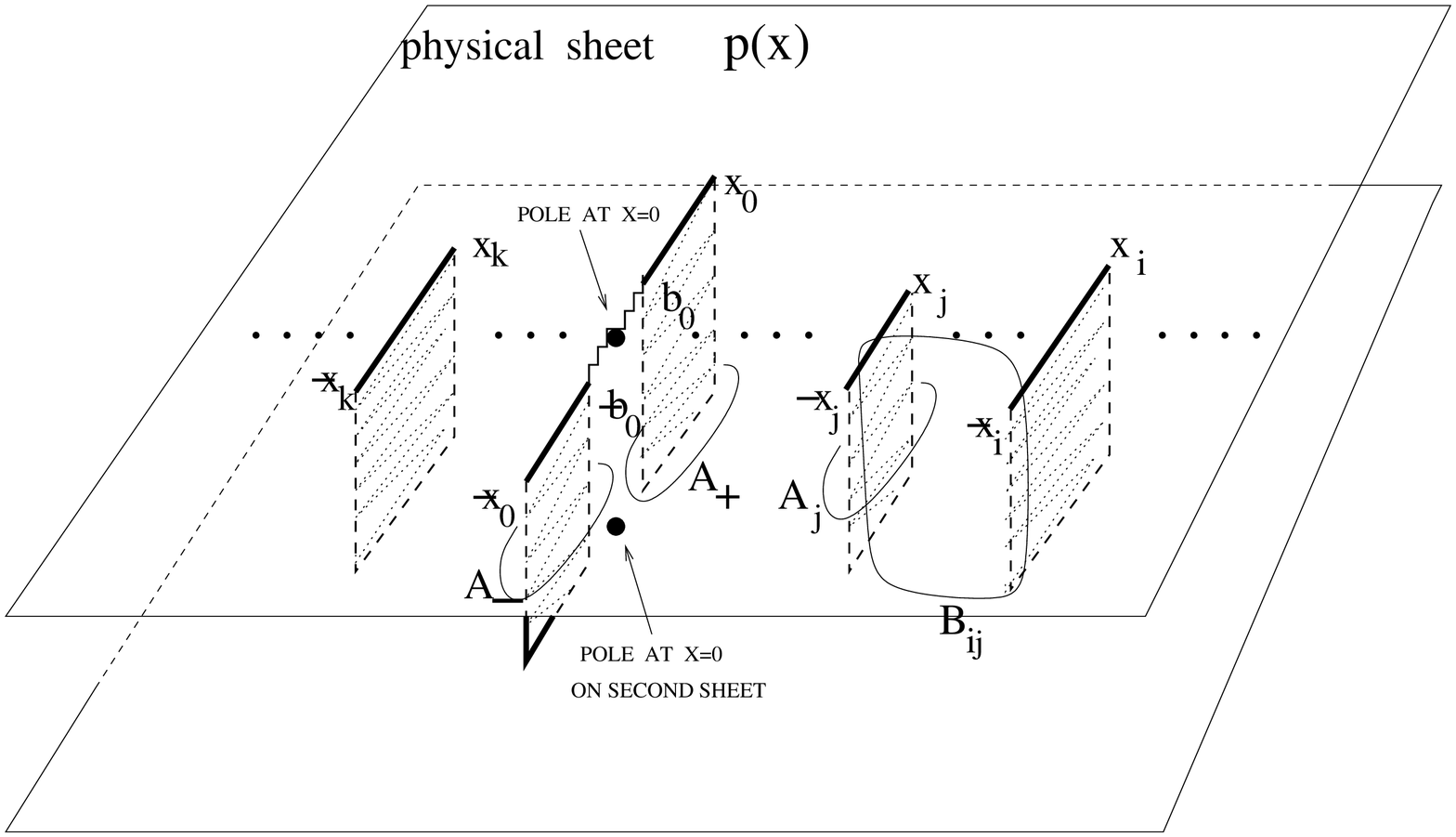,width=120mm}}
\caption{Hyperelliptic Riemann surface $\Sigma$, defined by
equation (\ref{sigmaxxx}); zig-zag line represents the condensate cut.
There is a double pole of differential $dG$ at the image of the
origin $x=0$ on the second sheet, while the differential $dp$ has double
poles at $x=0$ on both sheets. The sheet where $dG$ has no singularities is
called physical.}
\label{fi:RIEMANN}
\end{figure}

Equation (\ref{der})  imposes an integrality condition
on the ${\bf B}$-periods of $dp$,
\be
\label{Bint}
\oint_{B_i}dp = 2\pi (n_i-n_K)\qquad  i=1,\ldots,K-1,
\ee
for the canonical choice of the ${\bf B}$-cycles on the hyperelliptic
surface, $B_i\equiv B_{iK}$ (see fig.~\ref{fi:RIEMANN}).
In the absence of
condensates the single-valuedness of $p(x)$ on a physical sheet
requires in addition that ${\bf A}$-periods are zero,
\be
\label{Aint}
\oint_{A_i}dp = 0\qquad  i=1,\ldots,K-1,
\ee
for the canonical $A_i$ cycles on the hyperelliptic
surface (fig.~\ref{fi:RIEMANN}).

\begin{figure}[tp]
\centerline{\epsfig{file= 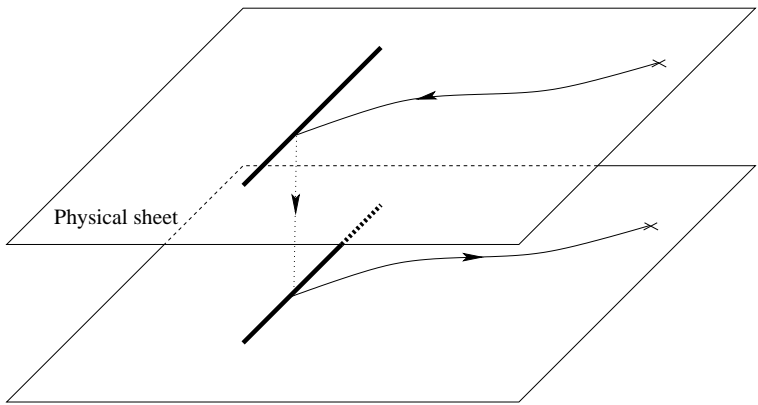,width=100mm}}
\caption{The contour of integration in \rf{ZEROMODE}. The marked points
on two sheets correspond to infinities $\infty_+$ and $\infty_-$ where
$x=\infty$.}
\label{openC}
\end{figure}

We need an extra condition to satisfy all $K$ equations
in (\ref{eqxxx}). To find it, one can rewrite the
$K$-th equation in (\ref{eqxxx}) as
\be
\label{ZEROMODE}
p(\infty_+)-p(\infty_-)=\int_{\infty_-}^{\infty_+} dp =2\pi n_K,
\ee
where by $\infty_+$ and $\infty_-$ we denoted two points with
$x=\infty$ on the first and second sheets of $\Sigma$ respectively.
The contour of integration comes from the infinity on the physical
sheet, passes through the $K$-th cut ${\bf C}_K$ onto the second
sheet, and then goes to the infinity on this sheet
(fig.~\ref{openC}). The points $\infty_+$ and $\infty_-$ are marked
by the fact that the differential $xdp$, which determines the number of Bethe
roots, has first order poles there (with residues of opposite sign),
though $dp$ itself is non-singular.  Note that since the choice of the
cut $K$ is arbitrary we could choose the set of equations similar to
\rf{ZEROMODE} on all $K$ cuts to replace the equations \rf{Bint}.

Since $dp$ has  second-order poles at $x= 0$ on both sheets
of the Riemann surface, it is of the general form
\be
\label{gsol}
dp =
{dx\over y}
 \sum_{k=-1}^{K-1}a_k x^{k-1}.
\ee
The pole at $x=0$ on the physical sheet must be cancelled in $G(x)$,
which requires $dp={dx\over 2x^2}+O(1)$ and fixes the first two
coefficients to be\footnote{We assume that the branch of the square root
is chosen such that $y=x^K+\ldots$ at infinity on the physical sheet.
The value of $y$ at zero may coincide with the algebraic square root
of $r_{2K}$ or may differ from it by a sign.
The square root $ \sqrt{r_{2K}}$ in this
formula should be
understood in the analytic sense as $y(0)$.
}
\be
a_{-1}=\frac{\sqrt{r_{2K}}}{2},
~~~~~a_0={r_{2K-1}\over
4\sqrt{r_{2K}}}.
\ee
The system of equations (\ref{Aint}) uniquely determines the rest of
the $K-1$ coefficients $a_k$ and thus completely defines the
differential (\ref{gsol}) on a given Riemann surface
(\ref{sigmaxxx}). The rest of the conditions are conditions on the
moduli of the Riemann surface itself.

All together the curve (\ref{sigmaxxx}) is parameterized by $2K$
end-points of
the cuts ${\bf C}_i$. The integrality of ${\bf B}$-periods expressed by the
eqs.(\ref{Bint}) and the extra condition (\ref{ZEROMODE})
constitute a system of $K$ equations on the
coefficients of (\ref{sigmaxxx}) and thus reduce the total number
of free parameters to $K$.  These free parameters exactly
correspond to the filling
fractions of the Bethe roots on the $K$ cuts ${\bf C}_j$,
\be
\label{frac}
S_j = {1\over 2\pi i}\oint_{A_j}p(x) dx =
\int_{{\bf C}_j} \rho(x)dx, \ \ \ \ j=1,\dots,K-1
\\
S_K = {1\over 2\pi i}\oint_{{\bf C}_K}p(x) dx =
\int_{{\bf C}_K} \rho(x)dx = \alpha - \sum_{j=1}^{K-1}S_j
\ee
The residue at infinity determines the total number of  Bethe roots
\footnote{Note, that (\ref{infty}) directly corresponds to the
asymptotics (\ref{asinf})
$$
dp= \left(\ha-\alpha\right) \,
\frac{dx}{x^2}+\ldots~~~(x\rightarrow\infty) $$}:
\be
\label{infty}
\frac{1}{2}-\alpha = -
{1\over 2\pi i}
\res_\infty  xdp
\ee
The normalization condition immediately allows us to set $a_{K-1}=\half-
\alpha$.
The total momentum (\ref{MOMrho}) can be rewritten  in terms
of the filling fractions $S_j$ in  (\ref{frac}) as
\be
\label{MOMG2}
P= -{1\over 2\pi i}\res_{x=0}\ {p(x)dx\over x} =
2\pi\sum_{j=1}^K n_j S_j
= 2\pi m.   
\ee

The anomalous dimension (\ref{gamacl}) can be now expressed directly
through the parameters of the hyperelliptic curve (\ref{sigmaxxx}) and
of the quasimomentum (\ref{gsol}) or resolvent $G(x)$.  From the expansion
of $p(x)$ at $x=0$, $$p(x)= -{1\over 2x}-2\pi m-{8\pi^2L\gamma\over
\lambda}\, x + O(x^2),$$ we have
\be
\label{gamma}
\gamma = { \lambda\over 8\pi^2L  }\left({r_{2K-2}\over 4 r_{2K}}
-\frac{r^2_{2K-1}}{16r^2_{2K}}
-{a_1\over \sqrt{r_{2K}}}\right)
\ee
which is a general solution for the one-loop anomalous dimension of
``long" operators \rf{typop}. It is expressed explicitly through the
coefficients of the equation \rf{sigmaxxx} which defines the complex curve,
and by $a_1$ which is determined through the solution of the linear system \rf{Aint}, and is
a transcendental function of the coefficients of \rf{sigmaxxx}. The coefficients
of \rf{sigmaxxx} are themselves defined implicitly as functions of the mode
numbers $n_j$, the total momentum $P=2\pi m$,
and the filling fractions $S_j$.

\begin{figure}[tp]
\centerline{\epsfig{file= 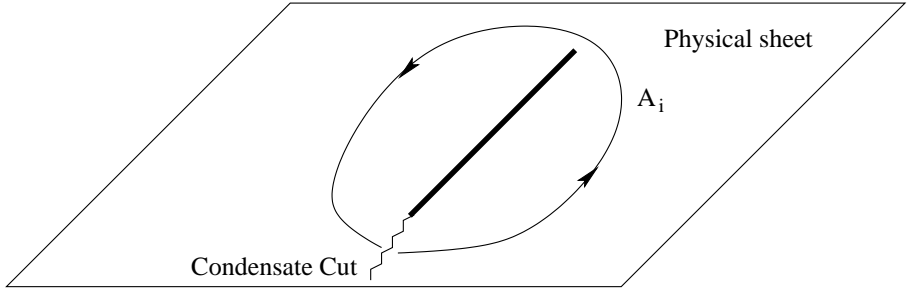,width=120mm}}
\caption{When an A-cycle crosses the condensate,
$\oint_{A_i}dp=\,{\rm Disc}\, p=2\pi m_i$.}
\label{fi:condensate}
\end{figure}

Very little changes if we allow for condensates. The differential
of the quasi-momentum is still an Abelian differential of the second
kind, the equations
\rf{eqxxx} are still solved by imposing integrality of ${\bf B}$-periods
on $dp$ and an extra condition \rf{ZEROMODE}.
The only difference is in the ${\bf A}$-periods. If an A-cycle $A_i$ crosses the
condensate cut, $p$ jumps by $2\pi m_i$, where
$m_i$ is an integer. Hence, ${\bf A}$-periods are no longer zero but are
 also integral (fig.~\ref{fi:condensate}). We thus
get a very symmetric set of conditions on the periods of $dp$,
\be\label{ABint}
\oint_{B_i}dp = 2\pi (n_i-n_K),\\
\oint_{A_i}dp = 2\pi m_i
\ee
 The counting of parameters does not change. There are
$2K-1$ equations (A and B cycle conditions and \rf{ZEROMODE})
on the total of $3K-1$ parameters ($2K$ moduli of the hyperelliptic
curve $\Sigma$ and $K-1$ free coefficients in the definition of
$dp$. eq.~\rf{gsol}). The $K$ parameter freedom reflects
the possibility to redistribute Bethe roots over the $K$ cuts
with arbitrary filling fractions \rf{frac}. The overall normalization
\rf{infty} fixes the total number of the roots and leaves $K-1$ independent 
parameters.

The approach of this section allows one to calculate the resolvent
and consequently all the conserved charges for a given Bethe state
in the scaling limit. It may seem that Bethe states
are in one-to-one correspondence with the sets of $K$ filling
fractions and $2K-1$ integers $\{n_i, m_i\}$, but this is not quite true.
Not all sets of parameters are allowed, since some of them
give rise to complex curves that do not satisfy
the reality condition, which therefore is a constraint on allowed
values of integers $n_i$ and $m_i$. Also, some different sets of $n$'s
and $m$'s may correspond to the same Bethe state. Suppose that we
solved for the resolvent and found all its branch points
${\rm x}_j$ as functions of the parameters. There are still many ways
to connect these branch points by cuts ${\bf C}_l$. In principle,
the cuts should be defined in such a way that the density
$\rho(x)dx$ for
$\rho(x)=\,{\rm Disc}\,G(x)={1\over 2\pi i}\left(G(x+i0)-G(x-i0)\right)$
is real and positive definite
on the cuts. Near each branch point $dx\,\rho(x)\sim
dx\,(x-{\rm x}_j)^{1/2}$. There are three lines on which
$dx\,(x-{\rm x}_j)^{1/2}$ is real and positive, and therefore three cuts
consistent with positivity can meet at each of the branch points.  This
leads to a discrete ambiguity in cutting the complex plane.  This
argument is standard in the discussion of multi-cut solutions of matrix
models (see, e.g. \cite{David:sk,Felder:2004uy}). Any rearrangement of cuts
or permutation of the branch points
induces a linear transformation of the ${\bf A}$- and ${\bf B}$- cycles
and therefore, according to
\rf{ABint}, of the integers $n_i$ and $m_i$. The branch of the square
root on the physical sheet may, however, change sign at infinity under
such rearrangement. The change in sign of the residue of $xdp$ at
infinity, according to \rf{infty}, is equivalent to interchanging
$\alpha$ with $1-\alpha$. If $\alpha >\half$ from the
outset, by rearranging cuts we can always get a new $\alpha$ in the
physical domain $0<\alpha<\half$. Therefore, an analytic
continuation of a solution of the Bethe equations beyond $\alpha=\half$
indeed corresponds to a physical Bethe state with $\alpha<\half$.  The
energy and all local charges are modular invariant and
do not change under the rearrangements of
cuts. Since local charges completely characterize a Bethe state, we
can relax even the reality condition for the density, and connect the
branch points by cuts in an arbitrary way. Then all $Sp(K-1,
{\Z})$ transformations of cycles are allowed. In particular, it is
always possible to set all ${\bf A}$-periods of $dp$ to zero by an appropriate
$Sp(K-1,
{\Z})$ transformation. In other words, it is always
possible to get rid of the condensates. This remark will be very
important in the next section, when we will compare general solutions
of the Bethe ansatz with classical solutions of the sigma-model. This
discussion may seem somewhat abstract, but we hope to clarify it with
explicit calculations for two-cut solutions in
section \ref{ss:examples}.

A similar ambiguity affects the total momentum \rf{MOMG2}
\be\label{tm}
P=-G(0)=\int_{0}^{\infty_+ }dG.
\ee
The contour of integration connects zero and infinity on the physical
sheet and is otherwise arbitrary. Adding an A-cycle $A_i$ to it shifts
the momentum by $2\pi m_i$, but the momentum is defined only up to an
integer multiple of $2\pi$, and only $ \e^{iP}$ makes sense.  The case
when the condensate passes through zero \cite{Beisert:2003xu} requires
special care, because in that case the integral in (\ref{tm}) is
ambiguous and should be defined symmetrically such that the
corresponding $A$-cycle contributes half of the period into the
momentum. With this prescription the condensate of density $m$ (Bethe
roots stretched along the imaginary axis with the spacing $im$) has
the momentum $m\pi$. Indeed, careful inspection of the general formula
(\ref{MOMu}) shows that an evenly spaced condensate has $\e^{iP}=1$,
while an oddly spaced distribution of roots gives $\e^{iP}=-1$.

The Bethe equations (\ref{BAE}) have an obvious symmetry,  $u_j \to -u_j$,
in which case the numerator replaces the denominator and vice versa.
Therefore, sometimes it is rather natural to consider a class of
solutions, when magnetic numbers satisfy
$n(-u_j)=-n(u_j)$,
Then the curve (\ref{sigmaxxx})
has a symmetry  $x \to -x$, i.e.
\be
\label{sigsym}
y^2 = \prod_{j=1}^K (x^2-{\rm x}_j^2)
\ee
The counting here is slightly different. We start from $K$ independent
parameters (instead of $2K$), impose $K/2$ constraints
(\ref{ZEROMODE}), since $K$ should be even in this case,
and the rest of parameters are eaten by the
residue of the differential $xdp$ (\ref{infty}) and half of the independent
``fractions"
(\ref{frac}). A particular example of a
symmetric solution will be considered in section \ref{ss:ellip}.

\subsection{ Two-loop corrections }

By this approach we can find also the two-loop corrections to our
general one-loop solution, thus diagonalizing the operator
\rf{dilatop} in the classical  limit, using the two-loop perturbative
integrability hypothesis proposed in
\cite{Beisert:2003tq} and confirmed in
\cite{Beisert:2003jb,Beisert:2003ys}.
A way to  take higher loop corrections
into account is to consider 
the Inozemtsev generalization of the spin-chain
in the hyperbolic limit \cite{Inozemtsev}, as was recently proposed 
by Serban and Staudacher
\cite{Serban:2004jf}.  

The Inozemtsev chain has long range interactions
between the sites, reflecting the fact that $g$-loop corrections
to the dilatation operator  involve spin-exchange separated
by up to $g$ sites.  The hyperbolic limit has one free parameter
that describes the fall-off of the interaction and Serban and Staudacher
\cite{Serban:2004jf}  
mapped this parameter to $\lambda/L^2$.  They then showed that not
only did
the Inozemtsev chain reproduce the two and three-loop dilatation
operators in \cite{Beisert:2003tq,Beisert:2003jb,Beisert:2003ys},
but it also matches the two-loop predictions for the string
solitons in \cite{Frolov:2003qc,Frolov:2003xy}.  However,
the three-loop predictions in \cite{Serban:2004jf} differ with those
in \cite{Frolov:2003qc,Frolov:2003xy}, and furthermore, there is
an explicit violation of BMN scaling starting at four loops
\cite{Serban:2004jf}.
Despite the fact that
the equivalence of the full perturbative SYM to the Inozemtsev
chain is still only a hypothesis, the
two-loop corrections are robust, and so we can use it
to study the anomalous dimensions of general operators at the two-loop
level and
compare these with predictions of quasiclassical string theory.

The two-loop corrected Bethe equations look
 as follows \cite{Serban:2004jf}
\be
\label{TWOLBA}
e^{ip_j L}=\prod_{k=1(k\ne j)}^J {u_j-u_k+i\over u_j-u_k-i},

\ee
where the momentum $p_j=p_j(u_j)$ is defined by solving the equation
\be
\label{DISPR}
u=u_0(p)+{\lambda\over 8\pi^2} \sin p +O(\lambda^2)
\ee
with the function $u_0(p)$ given by
\be\label{UJP}
e^{ip}={u_0+i/2\over u_0-i/2},
\ee
Up to two-loop order, eq.~\rf{TWOLBA} can be  rewritten as
\be
\label{CORRB}
\left({u_j+i/2-{\lambda\over 8\pi^2}{u_j\over u_j^2+1/4}
\over u_j-i/2-{\lambda\over 8\pi^2}{u_j\over
u_j^2+1/4}}\right)^L=
\prod_{k=1(k\ne j)}^J {u_j-u_k+i\over u_j-u_k-i},
\ee
This means that in the classical limit $u=xL$, $L\to\infty$,
eq. \rf{BAEC} is corrected as follows
\be
\label{BAETL}
{1\over x} +{\lambda\over 8\pi^2 L^2}{1\over x^3}   +2\pi n_l=
2 \pint_{{\bf C}} {d\xi\,\rho(\xi)\over x-\xi},\ \ \ x\in
{\bf C}_l
\ee
with the same normalization of the density \rf{RHON} as
in the one-loop case. Note however, that being perturbative from the gauge
theory side, the second term in the l.h.s. of \rf{BAETL} is ``singular"
perturbation of the integrable system, since it adds a singularity of higher
degree. For example, analogous perturbation in the algebraic matrix model
case \cite{David:sk,KM} would change the genus of the curve. Here, it will
causes some problems, when comparing with the string theory side.

The two-loop generalization of the total momentum quantization
condition
\rf{MOMu} can be rewritten as
\be
\label{QMQ}
\prod_{j=1}^J \left({u_j+i/2-{\lambda\over 8\pi^2}{u_j\over u_j^2+1/4}
\over u_j-i/2-{\lambda\over 8\pi^2}{u_j\over
u_j^2+1/4}}\right)^L= 1
\ee
or, in the classical limit,
\be
\label{QMQCL}
2\pi m=\int dx\,\rho(x)\left({1\over x}
+ {\lambda\over 8 \pi^2 L^2}{1\over x^3}\right) + O(\lambda^2)
 = - \int {dx\over 2\pi i}\,G(x)\left({1\over x}
+ {\lambda\over 8 \pi^2 L^2}{1\over x^3}\right) + O(\lambda^2)
\ee
The anomalus dimension is corrected as follows
\cite{Serban:2004jf}:
\be
\label{ANDIC}
\gamma=\sum_{j=1}^J\left[{\lambda\over 2\pi^2} \sin^2{p(u_j)\over 2}
-{\lambda^2\over 8\pi^4} \sin^4{p(u_j)\over 2}\right]+O(\lambda^3)=\\
=\sum_{j=1}^J\left[{\lambda\over 8\pi^2}{1\over
u_j^2+1/4}+{3\lambda^2\over 128\pi^4}{1\over
(u_j^2+1/4)^2}-{\lambda^2\over 128 \pi^4} {1\over
(u_j^2+1/4)^3}\right]+O(\lambda^3)
\ee
In the classical limit $L\to\infty$ this two-loop formula corrects
eq.\rf{gamacl} in the following way
\be
\label{GAMCOR}
\gamma=L\int dx\,\rho(x)\left({\lambda\over 8\pi^2L^2}{1\over x^2}
+ {3\lambda^2\over 128 \pi^4 L^4}{1\over x^4} +O\left({\lambda^2\over
L^6}\right) +O(\lambda^3)\right) =
\\
= - \oint {dx\over 2\pi i}\,G(x)\left({\lambda\over 8\pi^2L}{1\over x^2}
+ {3\lambda^2\over 128 \pi^4 L^3}{1\over x^4} +O\left({\lambda^2\over
L^5}\right) +O\left(\lambda^3\right)\right)
\ee
We kept here only the terms relevant in the BMN limit of fixed
$T={\lambda\over 16\pi^2 L^2}$.

We can now solve  eq. \rf{BAETL} in the same way as \rf{BAEC}.
We introduce
\be
\label{PDEF}
p(x)=G(x)-{1\over 2x}-{T\over x^3}
\ee
and find for the differential $dp$ on the Riemann surface
\rf{sigmaxxx} the following modification to \rf{gsol}
\be
\label{DPCORR}
dp={dx\over y}\sum_{k=-3}^{K-1} a_k x^{k-1}.
\ee
with the condition that $dp={1\over 2x^2}+{3T\over x^4}+ O(1)$ for
$x\to0$.  It is clear that this gives two extra  conditions
to fix the two extra  coefficients  $a_{-3}$ and $a_{-2}$.

Below we will compare these results with the two-loop correction
following from the sigma-model results, where the term ${1\over x^3}$
will appear from the two-pole structure $\ha\left({1\over x-1}+{1\over
x+1}\right)\simeq {1\over x}+{1\over x^3}+ O\left({1\over x^5}\right)$
in the integral equation similar to \rf{BAETL}.  We will see that
although the structures are very similar, the details are different.
In contrast to one loop, where the equations are easily reproduced
from those of the string sigma model, the two-loop equations of this
section follow from the integrable equations of string
theory only after a reparameterization of the spectral parameter.
This basically reflects the fact that the choice of the spectral
parameter
in \rf{DISPR} and \rf{BAETL} is somewhat arbitrary.
Nevertheless, they lead to the same
results for the anomalous dimension up to the two-loop level.
In section
\ref{ss:examples} we will present a wide class of examples where the
agreement is explicitly checked.

\sectiono{Classical integrability
\label{ss:classic}}

We now turn to integrability on the string side of the AdS/CFT
duality.  Classical integrability of the sigma-model in the
$AdS_5\times S^5$ background
\cite{Mandal:2002fs,Bena:2003wd} means, among other things,
that the classical equations of motion of a string in $AdS_5\times S^5$
can be solved by algebro-geometric methods
\cite{Pohlmeyer:1975nb,Zakharov:pp,book_of_soliton,DKN}
or, for certain particular cases, by the inverse scattering transformation
\cite{Faddeev's_book}. In this section,
we will examine finite-gap periodic solutions of the sigma-model and
establish direct links between them and the Bethe ansatz. A pedagogic
introduction into finite-gap integration methods can be found, for
example, in the monographs \cite{book_of_soliton,DKN}, and the
comparison of quantum inverse scattering with classical integrability,
applied mostly to the nonlinear Schr\"odinger and KdV equations, can be
found in
\cite{RESHSM,Smirnov:1998kv}.

As in the discussion of the SYM operators
we will focus on an $SU(2)$ reduction of the full sigma-model
 to the subsector of string moving on $S^3\times
{R}_t$. Many explicit
solutions are known in this case \cite{Tseytlin:2003ii}. The string
action in the conformal gauge is
\be
\label{sigmamo}
S_{\sigma m}
={\sqrt{\lambda}\over 4\pi }\int_0^{2\pi} d\sigma\, \int d\tau\,
\left[ \left(\d_aX_i\right)^2 -\left(\d_a X_0\right)^2\right],
\ee
where $X_0$ is the global AdS time and $X_i$, $i=1,\ldots, 4$ are
Cartesian coordinated on $S^3$ embedded in $
{R}^4$:
\be\label{sphere}
X_iX_i=1.
\ee
All other world-sheet coordinates (one radial and three angular coordinates
on $AdS_5$ and two extra angles on $S^5$) are set to constant values.
Such an ansatz describes a string localized in the
centre of AdS and moving in an $S^3$ subspace of $S^5$.
The effective
string tension is related to the 't~Hooft coupling according to AdS/CFT
\cite{Maldacena:1998re}. The equations of motion that follow from \rf{sigmamo}
should be supplemented with the Virasoro constraints:
\be\label{virasoro}
(\d_\pm X_i)^2=(\d_\pm X_0)^2 = \kappa^2,
\ee
where $\sigma_\pm=\half(\tau\pm\sigma)$, $\d_\pm=\d_\tau\pm\d_\sigma$,
and we always consider a gauge  $X_0=\kappa\tau$.

It was shown in \cite{Kruczenski:2003gt} that the equations of motion for
the string whose centre of mass moves along a big
circle of $S^3$ with a large angular momentum have
a well-defined weak-coupling limit ($\lambda/L^2)\ll 1$,
where now $\sqrt{\lambda}$
is the effective string tension and $L$ is the angular momentum,
and reduce in this limit to\footnote{Our normalization
of the world-sheet coordinates $\tau$ and $\sigma$ is different from
that of
\cite{Kruczenski:2003gt}.
We take $0\le\sigma\le 2\pi$ and rescale $\tau$ by the total spin $L$,
consistently with the normalization ${\bf S}^2=1$.}
\be\label{krucz}
\sin\theta\partial_\tau\theta+\partial_\sigma(
\sin^2\theta\partial_\sigma\varphi)=0
\\
\partial^2_\sigma\theta-\sin\theta\partial_\tau\varphi
-\sin\theta\cos\theta(\partial_\sigma\varphi)^2=0,
\ee
where $\varphi$ and $\theta$, in addition to another angle that
parameterizes the motion of the centre of mass, are
coordinates on $S^3$. This limit can be regarded as an approximation of
an almost light-like string motion \cite{Mikhailov:2003gq}
In terms of a unit vector,
${\bf S}=(\sin\theta\,
\cos\varphi,\sin\theta\,\sin\varphi,\cos\theta)$ the equations \rf{krucz}
take the form
\be
\label{geimag}
{\bf S}_\tau = {\bf S}\times {\bf S}_{\sigma\sigma}.
\ee
This is the equation of the classical Heisenberg model. It can be
formally obtained from operator equations of motion in the spin chain
by replacing quantum spins $\bsigma_l$ with ${\bf C}$-number
unit vectors ${\bf S}$ and simultaneously taking the
continuum limit: $\bsigma_l(\tau)
\rightarrow{\bf S}(2\pi l/L,\tau)={\bf S}(\sigma,\tau)$.
In \cite{Kruczenski:2003gt} Kruczenski
presented a path-integral derivation of this fact,
and identified this classical approximation
with the limit of the large size of the chain in which only slowly varying
spin configurations are kept.

The classical Heisenberg equation is completely integrable
 \cite{Lakshmanan,Takhtajan:rv,Zakharov:jc,Faddeev's_book}
and is a particular case of a more general
Landau-Lifshitz equation
\be
\label{LL}
{\bf S}_t = {\bf S}\times {\bf S}_{\sigma\sigma} + {\bf S}\times J{\bf S}
\ee
where $J$ is an anisotropy matrix which is proportional to the
unit matrix in our
case, so that second term in the r.h.s. vanishes and
(\ref{LL}) turns into (\ref{geimag})
\footnote{Note that the dynamical Neumann systems considered in
\cite{Tseytlin:2002ny}-\cite{Tseytlin:2003ii} can be  considered
as a particular reduction of equation \rf{geimag}, and we do not
necessarily need to introduce the anisotropy \rf{LL}, contrary to the
proposal of
\cite{Gorsky:2003nq}.}. We examine the classical Heisenberg
equations in the next section
and establish an equivalence between their periodic solutions and the
general scaling solution of the Bethe ansatz
which was obtained in section \ref{ss:generic}.

\subsection{Finite-gap solutions of the Heisenberg magnet}

In order to relate integrable structures of the classical and quantum
Heisenberg models, let us perform a continuous limit directly to
the transfer
matrix \rf{trama}. If we replace $\bsigma_l$ in \rf{trama} by
${\bf S}(2\pi l/L)$
and also rescale the spectral parameter $u=Lx$, then
the transfer matrix becomes
$$
\frac{T(Lx)}{(Lx)^L}\rightarrow{\rm Tr}\,\prod_{l=L}^{1}
\left(1+\frac{i}{2xL}\,{\bf S}(2\pi l/L)\otimes{\bsigma}\right).
$$
This expression
has a well-defined continuum limit,
as an anti-path-ordered exponential
\be
\label{clamo}
\prod_{l=L}^{1}
\left(1+\frac{i}{2xL}\,{\bf S}(2\pi l/L)\otimes{\bsigma}\right)\
\stackreb{L\to\infty}{\rightarrow}
\Omega(x)=\bar{P}\exp\left(\frac{i}{2x}\oint \frac{d\sigma}{2\pi}\,
{\bf S}(\sigma)\cdot{\bsigma}\right)
\ee
which is just the monodromy matrix of the classical Heisenberg model, and
we can introduce the classical quasimomentum $p(x)$ by
\be
\Tr\Omega(x) \equiv \lim_{L\rightarrow\infty}\frac{T(Lx)}{(Lx)^L}
= 2\cos p(x)
\ee
The classical equations of the Heisenberg magnet
are equivalent to the consistency condition $[L,M]=0$
of an auxiliary linear problem
\be
\label{auxili}
L\Psi = \left(\d_\sigma - \frac{iU}{4\pi x}\right)\Psi(x;\tau,\sigma) = 0
\\
M\Psi = \left(\d_\tau -\frac{iU}{4\pi^2x^2}+\frac{ iV}{4\pi x}\right)
\Psi(x;\tau,\sigma) = 0
\ee
where the Lax matrices are
\be
\label{UVS}
U={\bsigma}\cdot{\bf S}
\\
V={\bsigma}\cdot\left({\bf S}_\sigma\times{\bf S}\right)
\ee
The monodromy matrix is a parallel transporter of the
flat connection $(L,M)$ around the circle:
$\Omega(x) =\Psi(x;\tau_0,1)$, where $\Psi(x;\tau,\sigma)$ is the solution
of (\ref{auxili}) with the initial condition $\Psi(x;\tau_0,0)=1$.
Since the connection is flat
the trace of the monodromy matrix does not depend on $\tau_0$
and therefore generates an infinite set of integrals of motion.
The classical monodromy matrix is unimodular and unitary
when the spectral paramater is real. Its eigenvalues $\e^{\pm ip(x)}$
then determine the quasi-momentum $p(x)$,
and the Taylor expansion of the quasi-momentum generates an
infinite set of conserved
charges, which are local functionals of ${\bf S}(\tau,\sigma)$
\cite{Faddeev's_book}.

{}From section \ref{ss:bethe} we know that taking the scaling limit
of the quantum transfer matrix in \rf{eigenv} leads to
\be\label{classp}
2\cos p(x)\equiv\frac{T(Lx)}{(Lx)^L}=2\cos\left(G(x)-\frac{1}{2x}\right),
\ee
with the resolvent defined in \rf{defG}. There is  a $1/(2x)$ correction term
in (\ref{classp})
because, unlike in \rf{tu},  we did not shift the spectral parameter by $i/2$.
 We conclude that the resolvent constructed from
the density of Bethe roots and the quasi-momentum of the classical problem
are related exactly as in (\ref{pofG}).
The local conserved charges in the spin chain and in the classical
Heisenberg model are thus equivalent.  This observation elucidates the
meaning of the transformations used in \cite{Arutyunov:2003rg} to
relate higher conserved charges in the sigma-model to the conserved
charges of the spin chain. The approach of \cite{Arutyunov:2003rg} is
based on B\"acklund charges, which differ from charges defined by the
expansion of quasi-momentum. One has to do a linear transformation on
an infinite set of charges to account for this difference.

The transformation of the quantum spin operators into the c-number
unit vectors is hard to motivate rigorously in the present context.
There is a more direct way to compare the Bethe ansatz of the quantum spin
chain and the auxiliary linear problem of the classical model, which
allows one to derive \rf{pofG} directly, by inspecting the analytic
structure of the quasi-momentum.
Moreover, this method can be generalized to the sigma-model, where
the quantum counterpart (all-loop Bethe anzatz) is not yet known.

It is always possible to reconstruct the solution of the Heisenberg
equation from the solution of the auxiliary linear problem.
Indeed, it follows from \rf{auxili}, and \rf{UVS} that
\be\label{Spsipsi}
{\bf S}=-2\pi ix\Tr\left(\bsigma\,\Psi_\sigma\Psi^{-1}\right).
\ee
A slightly less trivial statement is that $\Psi(x;\tau,\sigma)$ is
essentially determined by the analytic properties of the quasi-momentum $p(x)$.
Let us now establish
these properties by inspecting the auxiliary problem \rf{auxili}.

\begin{figure}[tp]
\centerline{\epsfig{file=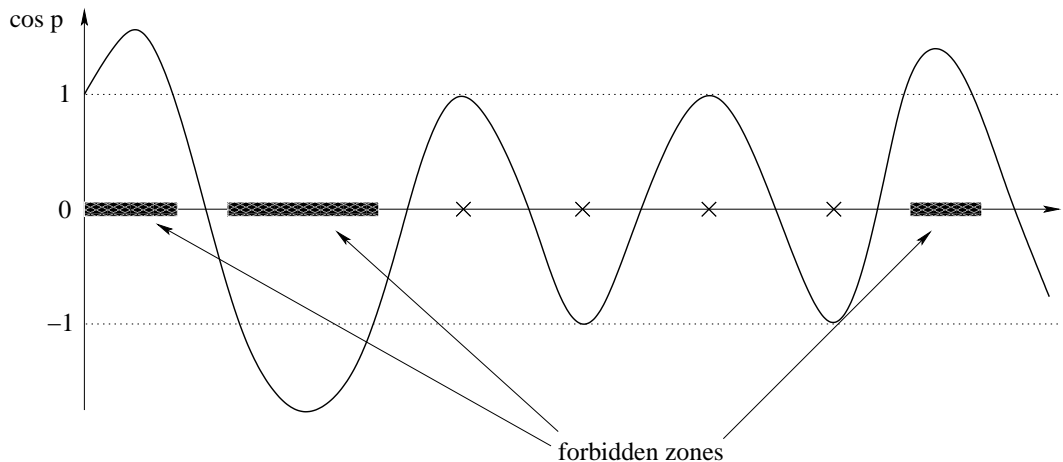,width=140mm}}
\caption{The trace of the monodromy matrix as a function of the
spectral parameter along the curve on which $\cos p(x)$
is real. The crosses correspond to degenerate periodic solutions,
which represent forbidden zones shrunk to points.}
\label{brillouin}
\end{figure}

The linear problem \rf{auxili}
has a form of a one-dimensional Dirac equation
\be\label{dirac}
\left(\d_\sigma - \frac{i{\bf S}\cdot{\bsigma}}{4\pi x}\right)\psi=0,
\ee
were $\psi$ is a two-component spinor.
 Two linearly independent solutions of this equation
can be chosen quasi-periodic. Indeed, if the initial
conditions are eigenvalues of the monodromy matrix
\be
\Omega(x)\psi_\pm(x;0)=\e^{\pm i p(x)}\psi_\pm(x;0),
\ee
the solution $\psi_\pm(x;\sigma)=\Psi(x;\sigma)\psi_\pm(x;0)$ will
again satisfy
$\psi_\pm(x;\sigma+2\pi)=\e^{\pm i p(x)}\psi_\pm(x;\sigma)$
because $\Psi(x;\sigma+2\pi)=\Psi(x;\sigma)\Omega(x)$.

However, the quasi-momentum
is not neccesarily real, being in general, a complex-valued
function of the spectral parameter. The condition that its imaginary part
vanish is a single equation on two variables (the real and imaginary
parts of the complex spectral parameter $x$).
Therefore, the quasi-momentum is real on a set of disjoint one-dimensional
supports (some segments in the complex plane of $x$). These
segments correspond to allowed
zones of the spectral problem \rf{dirac}. In a more familiar
setup of the  Schr\"odinger equation with a periodic potential,
the allowed energy zones
lie on the real axis, but the operator \rf{dirac}
is not Hermitean.
The allowed zones are singled out by the condition
that ${\rm Im}\,\left({\rm Tr}\,\Omega\right)=0$ and
$|{\rm Tr}\,\Omega|<2$. One can also define forbidden zones as loci where
${\rm Im}\,\left({\rm Tr}\,\Omega\right)=0$ and $|{\rm Tr}\,\Omega|>2$,
see fig.~\ref{brillouin}.
In general the number of allowed and forbidden zones is infinite.
We will only discuss finite-gap, or
algebro-geometric solutions for which this number is finite. They are
governed by the geometry of a complex curve of finite genus \rf{sigmaxxx}
and directly
correspond to solutions of the Bethe equations with a finite number of cuts.
In other words, it is more convenient to work with the smooth non-degenerate
curve \rf{sigmaxxx} than with the curve, defined by the Baxter equation
on quasimomentum $\det (\Omega - e^{ip}) =0$ or
\be\label{LAMDAOMEGA}
\e^{ip}+\e^{-ip} = \Tr\Omega(x)
\ee
which is generally a degenerate curve of an infinite genus\footnote{Note, that
here is an essential difference with the quasiclassical chain with
a  finite
number $L$ of large spins. In the latter case \rf{LAMDAOMEGA} basically
coincides with \rf{sigmaxxx} and the maximal number of cuts is bounded
by the length $L$ of the chain.
We restrict ourselves to this wide class of finite-gap or algebro-geometric
solutions, though
the equivalence between the Bethe ansatz and the classical solutions certainly
holds for infinite-zone solutions too.}.

At zone boundaries the monodromy matrix degenerates into the Jordan cell and
has only one eigenvector with an eigenvalue $1$ or $-1$. The quasi-momentum
becomes an integer multiple of $\pi$ and the two independent solutions of
\rf{dirac} become degenerate.
This degenerate solution of (\ref{dirac}) is either
periodic or anti-periodic. Accidental periodic
and anti-periodic solutions may appear inside some of the allowed zones
(see fig.~\ref{brillouin}). They are always present for finite-gap potentials,
and this is related to the fact that $p(x)$ is not an algebraic function
on \rf{sigmaxxx}.
The  linear problem (\ref{dirac}) still has two independent (anti)-periodic
solutions at these special points. Hence,
$\Omega$ is either $+{\bf 1}$ or $-{\bf 1}$, and
these points do not correspond to zone boundaries.
They rather represent forbidden zones shrunk to a point
and of course are not generic,
as much as finite-gap solutions are not generic in the space of
all solutions to the Heisenberg magnet equations.

The zone boundaries are singular points for
$\psi_\pm(x;\sigma)$ as functions of $x$, because $\psi_\pm$
collapses into one degenerate solution there.
 It is thus natural to regard $\psi_+$ and $\psi_-$ as
two branches of a single analytic function $\psi$ on a double cover of the
complex $x$ plane.
The zone boundaries are branch points at which the two branches collide.
Apart from an essential singularity at $x=0$,
$\psi$ is a holomorphic function on a hyperelliptic curve, two sheets
of which are glued together along the forbidden zones.
This is how the Riemann surface $\Sigma$, defined by
(\ref{sigmaxxx}), arises in the auxiliary problem of the classical
Heisenberg model.
The two eigenvalues of the monodromy matrix,
$\e^{\pm ip(x)}$ are also branches of a single meromorphic function on
 $\Sigma$. In fact, the trace of the monodromy matrix is an entire
function of $x$. There is no reason for it to be singular anywhere,
except at $x=0$ where the potential in
\rf{dirac} has a singularity. But solving \rf{LAMDAOMEGA}
for $p$ we will encounter
square root singularities when the discriminant  of the quadratic  equation
 \rf{LAMDAOMEGA}, with respect to
$\e^{ip}$, becomes zero. This is another way to
see why
the quasi-momentum has branch points.

The matrix solution of \rf{auxili}, $\Psi(x;\tau,\sigma)$,
is a meromorphic function on $\Sigma$ as well.
This latter function has the following asymptotic behavior:
\be
\label{baf}
\Psi(x;\tau,\sigma)\ \stackreb{x\to 0}{\sim}\
\exp\left[i\sigma_3\left({\sigma\over 4\pi x} + {\tau\over 4\pi ^2
x^2}\right)\right]\cdot
\Phi(x;\tau,\sigma)\cdot C
\ee
where $C$ is just a constant matrix (independent of $x$, $\tau$ and
$\sigma$), and
\be
\Phi(x;\tau,\sigma) = \phi_0(\tau,\sigma) + \sum_{k>0}\phi_k(\tau,\sigma)x^k
\ee
is regular at $x=0$.  An analytic matrix function (\ref{baf}) on the
hyperelliptic Riemann
surface \rf{sigmaxxx} with such asymptotics and $g+1$ extra poles, which
can be chosen independent of $\tau$ and $\sigma$,
is unique, they can be expressed in terms of the
Riemann theta-functions \cite{Bob} by the standard technique of
finite-gap integration.

Let us return to the analytic properties of the quasi-momentum.
Consider the values of the quasi-momentum $p(x+i0)$ and $p(x-i0)$
on the two sides of a forbidden zone. Shifting infinitesimally
the point $x-i0$ towards the forbidden zone, one can pass it through
the cut onto the second sheet. Thus $p(x+i0)$ and $p(x-i0)$ can
be regarded as two branches of a double-valued function $p(x)$
at the same point on the complex $x$-plane.
One can think of  $\e^{ip(x+i0)}$ and $\e^{ip(x-i0)}$ as
two independent
solutions of the chracteristic \rf{LAMDAOMEGA}
equation for the monodromy matrix.
Since $\Omega(x+i0)$ is unimodular,  $\e^{ip(x+i0)}\e^{ip(x-i0)}=1$, and
the quasi-momentum must satsify
\be
\label{RHPR}
p(x+i0)+p(x-i0)=2\pi n_k,~~~~~x\in {\bf C}_k
\ee
on each of the forbidden zones.  This equation represents a
Riemann-Hilbert problem equivalent to (\ref{eqxxx}), if the
quasi-momentum and the resolvent are related as in \rf{pofG}.  The
integer $n_k-n_{k-1}-1$ is the number of (anti)-periodic solutions
within the $k$-th allowed zone, that is the number of times the
equation ${\rm Tr}\,\Omega(x)=2$ is satisfied as $x$ moves along the
allowed zone between the forbidden zones ${\bf C}_{k-1}$ and  ${\bf C}_k$.
Equivalently, $n_k-n_{k-1}-1$ is the number of oscillations $\cos
p(x)$ makes between the end-points of the forbidden zones ${\bf
C}_{k-1}$ and ${\bf C}_k$ (see fig.~\ref{brillouin}).

We see that, in the classical system, the integrality of
${\bf B}$-periods (\ref{Bint}) essentially follows from the unimodularity
of the transfer matrix. Since $p(x)$ is single-valued outside of the cuts,
all ${\bf A}$-periods are zero. In other words, there are no condensate cuts.
As we remarked in section \ref{ss:generic}, it is always possible to choose
cuts in such a way that condensates ``evaporate'' and all
${\bf A}$-periods of $dp$
 identially turn to zero. Such a choice of  cuts is unique and
corresponds precisely to cutting the $x$-plane along the forbidden zones.
The condition that defines the forbidden zones,
${\rm Im}\,{\rm Tr}\,\Omega=0$ and $|{\rm Tr}\,\Omega|>2$,
does not necessarily coincide with the condition that $-idx\,{\rm Disc}\,p$
be real and positive, which is  appropriate for the Bethe ansatz. This is
a crucial difference  as compared to
 the quantum KdV equation, where Bethe roots
condense on the forbidden zones of the
classical problem \cite{Smirnov:1998kv}.

A simple asymptotic analysis of the linear problem (\ref{dirac}),
(\ref{baf})
shows that the quasi-momentum
behaves as $p(x)\sim \pm {1\over 2x}~~~{\rm at}~ ~~x\rightarrow 0$,
this analysis amounts to solving equation

(\ref{dirac}) in the WKB approximation. Hence,
$G(x)=p(x)+{1\over 2x}$ is regular at zero on one of the sheets
of the Riemann surface (\ref{sigmaxxx}), and this singles out the physical
sheet of $\Sigma$.

To complete a proof of equivalence between the Bethe
equations and the classical auxiliary problem,
we should also check that the reality of ${\bf S}(\tau,\sigma)$
reproduces the reality condition on the distribution of
Bethe roots. The reality of ${\bf S}(t,\sigma)$ implies the following
conjugation property of the monodromy matrix:
\be
\overline{\Psi}(x,t,\sigma)=
\sigma_2\Psi(\bar{x};t,\sigma)\sigma_2,
~~~~~\overline{\Omega}(x)=\sigma_2\Omega(\bar{x})\sigma_2,
\ee
where the bar over matrices
denotes complex conjugation of all matrix elements. If
$\Omega(x)\psi=\e^{ip(x)}\psi$, then
$\overline{\Omega}(x)\bar{\psi}=\e^{-i\bar{p}(x)}\bar{\psi}$,
and
$\Omega(\bar{x})\sigma_2\bar{\psi}=\e^{-i\bar{p}(x)}\sigma_2\bar{\psi}$.
Therefore, $\sigma_2\bar{\psi}$ is an eigenvector of $\Omega(\bar{x})$
with an eigenvalue $\e^{-i\bar{p}(x)}$, i.e. $p(\bar{x})=\bar{p}(x)$,
which is indeed equivalent to the reality condition for the resolvent.

\subsection{Finite-gap solutions of the chiral field}
\label{ss:chiral}

Since the three-sphere $S^3$ is the group
manifold of $SU(2)$, the sigma-model (\ref{sigmamo})-\rf{virasoro} can be
equivalently reformulated as an $SU(2)$ principal chiral field:
\be
\label{su2}
g = \left(
\begin{array}{cc}
  X_1+iX_2 & X_3+iX_4 \\
 -X_3+iX_4 & X_1-iX_2
\end{array}
\right) \equiv
\left(
\begin{array}{cc}
  Z_1 & Z_2 \\
  -{\bar Z}_2 & {\bar Z}_1
\end{array}\right) \in SU(2),
\\
g^{-1}dg = \left(
\begin{array}{cc}
  {\bar Z}_1dZ_1 + Z_2d{\bar Z}_2 &  {\bar Z}_1dZ_2 - Z_2d{\bar Z}_1 \\
   {\bar Z}_2dZ_1 - Z_1d{\bar Z}_2 &  {\bar Z}_2dZ_2 + Z_1d{\bar Z}_1
\end{array}\right) \in su(2).
\ee
The action \rf{sigmamo} takes the form
\be
\label{cf}
S_{\sigma m}= -\frac{\sqrt{\lambda}}{4\pi}
\int d\sigma d\tau\, \left[\frac{1}{2}\,{\rm Tr} j_a^2+(\d_a X_0)^2\right].
\ee
where $j_a=g^{-1}\d_a g={1\over 2i}j^A_a\sigma^A$ with
$a=0,1=\tau,\sigma$, are the right currents.
The equations of motion can be
written in terms of their light-cone components as
\be\label{sm/eq}
\d_+j_-+\d_-j_+=0, \ \ \ \ \  \d_+\d_- X_0=0,
\\
\d_+j_--\d_-j_++[j_+,j_-]=0.
\ee
where the last equality follows from the expression for
the  currents in terms of the
matrices $g$ in \rf{su2}.

The sigma-model  on $S^3$  possesses a global
$SU_L(2)\times SU_R(2)$ symmetry which acts as
$g(\tau,\sigma)$ by left and right
multiplication by constant group matrices. This symmetry can be identified
with a particular $SO(4)$ subgroup of the $SO(6)$ $R$-symmetry group
of \4N SYM. The six scalar fields in the SYM  transform under $SO(6)$
in the same way as the
Cartesian coordinates on the five-sphere in the $AdS_5\times S^5$ geometry.
The  fields $\bPhi_1 $ and $\bPhi_2 $ have the same quantum numbers under
$SU_L(2)\times SU_R(2)$
as the sigma-model target space coordinates $Z_1$ and $Z_2$
defined in \rf{su2}, i.e.
$\bPhi_1 \sim Z_1$ and $\bPhi_2 \sim Z_2$.
Under the left shifts, generated by
\be
Q_L^A=\frac{\sqrt{\lambda}}{4\pi}\int d\sigma\, l^A_0,
\ee
where $l_a=gj_ag^{-1}=\partial_a g \,g^{-1}$, we have
$g\rightarrow hg$, and $(Z_1,-\bar{Z}_2)$ and
$(Z_2,-\bar{Z}_1)$ transform as doublets.
The normalization of the generator is such that $Z_1$ and $Z_2$
have $Q_L^3=1$. This implies that for an operator ${\rm
Tr}(\bPhi_1^{L-J}\bPhi_2 ^J+\ldots)$:
\be
Q_L^3=L.
\ee
Under the right shifts, generated by
\be
Q_R^A=\frac{\sqrt{\lambda}}{4\pi}\int d\sigma\, j^A_0,
\ee
$g\rightarrow gh$, and $(Z_1,Z_2)$ transforms as a
doublet, so that
$Z_1$ has  $Q_R^3=1$ and $Z_2$ has $Q_R^3=-1$.
Consequently,
\be
Q_R^3=L-2J.
\ee
The scaling dimension of an operator is dual to the energy
of the string solution,
\be
\Delta=\frac{\sqrt{\lambda}}{2\pi}\int_0^{2\pi} d\sigma\, \d_\tau X_0
=\sqrt{\lambda}\,\kappa,
\ee
which is generated by the global time translations.
Thus,
the Virasoro constraints \rf{virasoro} become
\be
\label{vircf}
\ha{\rm Tr} j_+^2=\ha{\rm Tr} j_-^2=-\kappa^2\,.
\ee
These constraints describe a consistent reduction of the $SU(2)$
chiral field introduced in \cite{Faddeev:1985qu}.
Upon the identification
\be
\label{jspm}
j_+=i\kappa\,{\bf S}_+\cdot{\bsigma},
~~~~~j_-=i\kappa\,{\bf S}_{-}\cdot{\bsigma},
\ee
the model reduces to a system of {\it two} interacting spins of unit
length ${\bf S}_+$ and ${\bf S}_{-}$:
\be
\label{emcf}
\d_+{\bf S}_{-}+2\kappa{\bf S}_{-}\times{\bf S}_{+}=0,
\\
\d_-{\bf S}_{+}-2\kappa{\bf S}_{-}\times{\bf S}_{+}=0.
\ee
These equations automatically incorporate the Virasoro constraints. The
model has an interesting quantum version
\cite{Faddeev:1985qu,Destri:1987hc,Faddeev:1996iy}, which is a spin
system with two spins per site. We do not see an obvious way to relate
the two-spin chain of \cite{Faddeev:1985qu} to the dilatation operator
of ${\cal N}=4$ SYM theory (though the Heisenberg model can be obtained in
a certain limit, see below), but do not exclude the possibility
that such a relationship may exist.

The equations of motion for the principal chiral field
can be written as a flatness condition \cite{Zakharov:pp}.
Define a current $J(x)$ dependent on a spectral parameter $x$ as
\be
J_\pm(x) = \frac{j_\pm}{1\mp x}\,.
\ee
Then the zero-curvature equation
\be\label{zerocur}
\d_+J_--\d_-J_++[J_+,J_-]=0
\ee
is equivalent to \rf{sm/eq}.

On the other hand, it is straightforward to show
that  \rf{zerocur} is
equivalent to the consistency condition of the following linear problem
\be\label{laxcf}
{\cal L}\Psi =
\left(\d_\sigma + \ha\left({j_+\over 1-x} - {j_-\over 1+x}\right)\right)\Psi
=\left[\d_\sigma + \frac{i\kappa}{2}
\left({{\bf S}_{+}\cdot{\bsigma}\over 1-x}
 - {{\bf S}_{-}\cdot{\bsigma}\over 1+x}\right)\right]\Psi
=0,
\\
{\cal M}\Psi =
\left(\d_\tau + \ha\left({j_+\over 1-x} + {j_-\over 1+x}\right)\right)\Psi
=\left[\d_\tau +
\frac{i\kappa}{2}\left({{\bf S}_{+}\cdot{\bsigma}\over 1-x}
 + {{\bf S}_{-}\cdot{\bsigma}\over 1+x}\right)\right]\Psi =0.
\ee
The solution to the linear problem  now has essential singularities
at $x \to \pm 1$, where near the singularity it has
a solution similar to \rf{baf}, and the
solutions to the chiral field equations \rf{emcf} are then defined as in
\rf{Spsipsi}. We are not going to discuss this issue here in detail, since
for comparison with section \ref{ss:bethe} we only need the spectral data of
the problem \rf{laxcf}.

Moreover, a more effective way to construct the
finite-gap solutions to sigma models was proposed in \cite{Krisig}, which
is generic for any number of fields in the sigma-model. It uses the
functions on
the double cover of the spectral curve \rf{sigmaxxx} and its generalizations
for larger number of fields. This construction is reviewed in
Appendix~\ref{ss:kri}.

The first equation of \rf{laxcf} defines the monodromy matrix:
\be
\Omega(x)=\bar{P}\exp\int_0^{2\pi} d\sigma\,
\ha\left( {j_+\over 1-x} - {j_-\over 1+x}\right),
\ee
with the quasi-momentum:
\be
\Tr\Omega(x)=2\cos p(x).
\ee
The rest of the story does not differ much from the analysis of the
spectral problem for the classical Heisenberg model. The only
difference is that the Lax pair \rf{laxcf} now has two poles
at $x=\pm 1$ instead of one.
The standard asymptotic analysis yields
\be\label{pres1}
p(x)=-\frac{\pi\kappa}{x\pm 1}+\ldots
~~~~~(x\rightarrow\mp 1).
\ee
To express the charges in terms of the spectral data, we expand the
quasi-momentum at
zero and at infinity. At infinity,
$\LL=\partial_\sigma+j_0/x+\ldots$, and
\be
\Tr\Omega=2+\frac{1}{2x^2}\int_0^{2\pi}d\sigma_1 d\sigma_2\,
\Tr j_0(\sigma_1) j_0(\sigma_2)+\ldots
=2-\frac{4\pi^2 Q_R^2}{\lambda x^2}+\ldots
=2-\frac{4\pi^2 (L-2J)^2}{\lambda x^2}+\ldots
\,,
\ee
Here we assume that
the classical solutions describe highest-weight
$SU_L(2)\times SU_R(2)$ states. For highest weights in large representations,
 $|Q_{L,R}|^2$ can be replaced with
$(Q_{L}^3)^2=L^2$ and $(Q_{R}^3)^2=(L-2J)^2$.
Thus we have
\be
p(x)=-\frac{2\pi(L-2J)}{\sqrt{\lambda}\, x}+\ldots~~~~~(x\rightarrow\infty).
\ee
At $x\to 0$, $\LL=\partial_\sigma+j_1-x j_0+\ldots$, which can be
written as $\LL=g^{-1}(\partial_\sigma-x l_0+\ldots)g$.
Then,
$$
\Omega(x)=g^{-1}(2\pi)\overline{P}
\exp\left(-x\int_0^{2\pi}d\sigma\,l_0+\ldots\right)g(0).
$$
Because of the periodicity of
  $g(\sigma)$, $\Omega(0)=1$ and $p(0)=2\pi m$.
Expanding further, we get
\be
\Tr\Omega=2+\frac{x^2}{2}\int_0^{2\pi}d\sigma_1 d\sigma_2\,
\Tr l_0(\sigma_1) l_0(\sigma_2)+\ldots=2-\frac{4\pi^2Q_L^2}{\lambda }\,x^2
+\ldots=2-\frac{4\pi^2L^2}{\lambda }\,x^2
+\ldots\,.
\ee
Hence,
\be
p(x)=2\pi m+\frac{2\pi L}{\sqrt{\lambda} }\, x+\ldots,~~~~~(x\rightarrow 0).
\ee
By the same arguments as in the previous section,
the quasi-momentum is a meromorphic function on the complex
plane with cuts, satisfying equation (\ref{eqxxx}) on each cut.
Subtracting the singularities at $x\to\pm 1$, we get the resolvent
\be
G(x)=p(x)+\frac{\pi\kappa}{x-1}+\frac{\pi\kappa}{x+1},
\ee
Because the poles at $x=\pm 1$ cancel,
the resolvent is an analytic function on the physical sheet and, as
in \rf{defG}, it can be represented
as an integral of a positive density
\be\label{spectrepr}
G(x)=\int  d\xi\,\frac{\rho(\xi)}{x-\xi}\,.
\ee
The proof of this spectral representation is the strandard
argument based on analyticity of $G(x)$.
We define $2i\pi\rho(x)=G(x+i0)-G(x-i0)$. Then the right-hand
side can be represented by a contour integral with the contour
surrounding  all the cuts. The only singularity
of the integrand on the outside of the contour
is a pole at  $y=x$ with residue $G(x)$.
Shrinking the contour, we get \rf{spectrepr}.

The asymptotic behavior of the resolvent,
$G(x)\sim 2\pi[\kappa-(L-2J)/\sqrt{\lambda }]/x$,
 translates into the
normalization condition for the density:
\be\label{nn}
\int dx\,\rho(x)   
=\frac{2\pi}{\sqrt{\lambda}}(\Delta+2J-L).
\ee
The density is subject to two extra constraints, which follow
from the asymptotic behavior of the resolvent at zero:
\be\label{cbethemom}
-{1\over 2\pi i}\oint {G(x)dx\over x}=
\int dx\,\frac{\rho(x)}{x}=2\pi m
\ee
and
\be\label{nn1}
-{1\over 2\pi i}\oint {G(x)dx\over x^2}=
\int dx\,\frac{\rho(x)}{x^2}  
=\frac{2\pi}{\sqrt{\lambda}}(\Delta-L).
\ee
The quasi-momentum satisfies  \rf{RHPR}:
\be
p(x+i0)+p(x-i0)=2\pi n_k,~~~~~x\in {\bf C}_k
\ee
This equation is a simple consequence of unimodularity
of the transfer matrix. Taking into account the spectral representation
\rf{spectrepr}, it can be recast into
 an integral equation for the density:
\be\label{cbethe}
G(x+i0)+G(x-i0) =
2\pint d\xi\,\frac{\rho(\xi)}{x-\xi}=\frac{2\pi\kappa}{x-1}
+\frac{2\pi\kappa}{x+1}+2\pi n_k, \qquad x\in {\bf C}_k.
\ee
We obtained again the same Riemann-Hilbert problem as for the
long spin chain \rf{eqxxx} or for the classical Hisenberg ferromagnetic
\rf{RHPR}, but with a different pole-structure.
This equation is a direct generalization of \rf{BAEC} and can be
called the classical Bethe equation for the chiral field.  It can be
solved in the same way as
\rf{BAEC}.  Namely, we look for a differential $dp$ defined on the
hyperelliptic surface \rf{sigmaxxx}, having double poles: $dp\sim
dx \left[{\pi\kappa\over (x\pm 1)^2}+O\left(1\right)\right]$
at $x=\pm 1$, and behaving as $dp\sim {2\pi\over
\sqrt{\lambda}}(L-2J){dx\over x^2}$ at
$x\to\infty$. We write, generalising the eq. \rf{gsol}
\be\label{twopole}
dp=\pi\kappa{dx\over y}\left({y_+\over
(x-1)^2}+{y_-\over (x+1)^2 }+ {y'_+\over
x-1}+{y'_-\over x+1 }+\sum_{k=1}^{K-1} b_k x^{k-1}\right)
\ee
where $y_\pm=\left.y\right|_{x=\pm 1}$,
$y'_\pm=\left.{dy\over dx}\right|_{x=\pm 1}$, and the coefficients $b_k$ are
determined, as in \rf{Aint} by vanishing of $\bf A$-periods.

The challenge is to guess what quantum Bethe equations can reproduce
\rf{cbethe} and its solution \rf{twopole} in the scaling
limit.  This does not look like an easy problem.  The Bethe equations
for the Faddeev-Reshetikhin model \cite{Faddeev:1985qu} is an obvious
candidate, and not surprisingly -- this model was constructed as a
quantization of the chiral field. The Bethe equations of
 \cite{Faddeev:1985qu} indeed reduce to
\rf{cbethe} in the scaling limit. The parameters, however, do not
match literally.

\subsection{ Comparison of string theory to perturbative gauge theory}

We are now in a position to compare the results of the
sect.~\ref{ss:bethe} for the spin chain, that describes the one- and
two- loop perturbation theory of the \4N planar SYM gauge theory, with
the sigma model of the dual string theory.  We should compare either
their generals solutions
\rf{DPCORR} and \rf{twopole}, or the general
equations \rf{BAETL} and \rf{cbethe}, together with the asymptotic
conditions for the
resolvent $G(x)$ or quasi-momentum $p(x)$.

First we will try to get a qualitative idea at one loop, by comparing
the small $\lambda/L$ limit of sigma model to the classical magnet
introduced in this section as a naive continuous limit of the XXX spin
chain.

The general solution of the chiral field
should reproduce the solution of the spin chain in the
weak-coupling limit.  On the classical level one can take the limit
directly.
If the 't~Hooft coupling is small, then $\kappa=\Delta/\sqrt{\lambda}$
is large
and we can expand the solution of \rf{emcf} in the inverse
powers of this parameter:
\be\label{suv}
{\bf S}_\pm = {\bf S}\pm\frac{1}{4\kappa}\,{\bf S}\times{\bf S}_\sigma
-\frac{{\bf S}_\sigma^2}{32\kappa^2}\,\,{\bf S}
+O\left(\frac{1}{\kappa^3}\right).
\ee
The $O(1/\kappa^2)$ term ensures the correct normalization
of the classical spins:
${\bf S}_\pm^2=1$. The world-sheet time variable should also be
rescaled such that  $\d_\tau\to \d_\tau/(4\kappa)$. Then
$$
\d_\pm=\frac{1}{4\kappa}\,\d_\tau\pm \d_\sigma,
$$
~~~
$$
{\bf S}_-\times {\bf S}_+=-\frac{1}{2\kappa}\,{\bf S}_\sigma
+O\left(\frac{1}{\kappa^3}\right)
$$
and  the leading
$O(1)$ term in the equations \rf{emcf} cancels identically.
The next $O(1/\kappa)$
term gives the equation of the Heisenberg magnet:
\be
{\bf S}_\tau={\bf S}\times{\bf S}_{\sigma\sigma}.
\ee
This procedure is in fact equivalent to the derivation
of the Heisenberg equations from the sigma-model of
ref.~\cite{Kruczenski:2003gt}, which is based
on separation of fast and slow modes of the string and also
involves rescaling of the world-sheet time
\footnote{We are grateful to A.Tseytlin for discussions of these issues
and the subtleties concerning the general two-loop comparison,
as well as sharing with us unpublished results on a similar approach
to relating sigma-models with classical spin systems.}.

Plugging \rf{suv}  into  (\ref{laxcf}),
rescaling $x\rightarrow 4\pi\kappa x$ and taking $\kappa$
to infinity
\be
\label{laxxxx}
{\cal L} \rightarrow \d_\sigma - \frac{i{\bf S}\cdot\bsigma}{4\pi x}
+O\left(\frac{1}{\kappa}\right),
\\
4\kappa{\cal M} \rightarrow \d_\tau -
 \frac{i{\bf S}\cdot\bsigma}{4\pi^2 x^2}
-\frac{i({\bf S}\times{\bf S}_{\sigma})\cdot\bsigma}{4\pi x}
+O\left(\frac{1}{\kappa}\right),
\ee
we recover the Lax pair of the magnetic. It means that on level of
classical integrability, some finite-gap solutions of sigma-model on
string side turn into the classical solutions of the Heisenberg spin
chain.

However, to make quantitative comparison between the prediction of SYM
and string theory for the anomalous dimensions \rf{GAMCOR} and \rf{nn1}
up to two loops, we have to compare the solutions to
basic integral equations
\rf{BAETL} and \rf{cbethe}, together with the normalizations of
densities and zero-momentum conditions. In other words, we have to
see how the geometric data depends on the integrals of motion. In
order to do this, let us take directly the weak-coupling limit in the
integral equation \rf{cbethe}. Rescaling for example $x$ by a factor
of $4\pi L/\sqrt{\lambda}$, the normalization conditions \rf{nn},
\rf{cbethemom} and \rf{nn1} become
\be
\label{compa}
\int dx\,\rho (x)=\frac{\Delta}{2L}-\frac{J_1-J_2}{2L}
=\frac{J}{L}+\frac{\Delta-L}{2L }\,,
\\
-\oint {dx\over 2\pi ix}\,G(x) = \int dx\,\frac{\rho (x)}{x}=2\pi m,
\\
- \frac{\lambda }{8\pi ^2L}\oint {dx\over 2\pi ix^2}\,G(x) =
\frac{\lambda }{8\pi ^2L}\int  dx\,\frac{\rho (x)}{x^2}=\Delta -L.
\ee
and the integral equation \rf{cbethe} transforms into
\be\label{cbethe1}
2\pint d\xi\,\frac{\rho(\xi)}{x-\xi}=\frac{x\Delta/L}{x^2-\frac{\lambda }{16\pi ^2L ^2}}
+2\pi n_k, \qquad x\in {\bf C}_k.
\ee
In the limit $\lambda/L^2\rightarrow 0$, $\Delta =L+O(\lambda/L )$,
and these formulas exactly reproduce the scaling limit of Bethe equations
\rf{BAEC} for the one loop perturbative SYM,
together with the relations \rf{asinf},\rf{MOMx} and \rf{gamacl}
completely defining the anomalous dimension. We see that the
one-loop physics (at least in this $SU(2)$ sector) of the SYM is
immediately reproduced from the string sigma-model for very general
solutions.

Comparison to two loops takes a little more work.
If we expand the rhs of (\ref{cbethe1}) in powers of $T=\frac{\lambda}{16\pi^2L^2}$,
we see that the coefficient of the $1/x^3$ term is half that of
the corresponding coefficient in (\ref{BAETL}).  This suggests that
one should replace the spectral parameter in (\ref{BAETL}) with
$\tx=x+T/x+O(T^2)$. The string density is thus related to the gauge theory density
by a change of variables, $\rho_g(x)=\rho(x-T/x+\ldots)$, where
the subscript $g$ refers to the gauge theory solution in section 3.2.
To reproduce the results of section 3.2, we need to change
variables in the integral equation and in the normalization conditions.
The change of variables in the momentum condition in \rf{compa} gives
\be
\int dx\,\frac{\rho_g(x)}{x}\left(1+\frac{2T}{x^2}+\ldots \right)=2\pi m,
\ee
which is the same as \rf{QMQCL}. The energy conditions transforms to
\be\label{perturbativeenergy}
\Delta-L=2TL\int dx\,\rho_g(x)\left(\frac{1}{x^2}+\frac{3T}{x^4}+\ldots \right),
\ee
in agreement with \rf{GAMCOR}. The normalization of the density becomes
\be
\int dx\,\rho_g(x)\left(1+\frac{T}{x^2}\right)
=\frac{J}{L}+\frac{\Delta-L}{2L }\,.
\ee
The unwanted terms cancel in virtue of \rf{perturbativeenergy} in agreement
with the canonical normalization of the gauge theory density \rf{RHON}.
Finally,
$$
\pint\frac{dy\,\rho (y)}{x-\frac{T}{x}-y}=\pint\frac{dz\,\rho_g(z)}{x-z}
+\frac{T}{x}\int dz\,\frac{\rho_g (z)}{z^2}+\ldots\,,
$$
Using  the energy condition \rf{perturbativeenergy},
we find that $\rho _g(x)$ satisfies \rf{BAETL} with the two-loop accuracy.
We see that the sigma model matches the gauge theory prediction up to
two loops for all long XXX operators. The gauge theory resolvent is related
to the resovent in the sigma model in a simple way.
At one loop they are the same. At two loops, an $O(T)$ correction arises
because of the change of variables:
\be
G_g(x)=G(x)-T\frac{G'(x)-G'(0)}{x}+\ldots\,.
\ee
This equation describes the map of higher charges in the sigma model
to those in the gauge theory.

\sectiono{Examples
\label{ss:examples}}

Let us now turn to particular examples of our generic
solution.  We consider first the BMN case, which from integrable system
point of view corresponds to a solitonic degeneration of our
general finite-gap solution. Then we discuss in detail the solutions
with one and two cuts.

\subsection{BMN states
\label{ss:bmn}}

Let us show how the BMN formula for the  anomalous dimension
\cite{Berenstein:2002jq}
is  reproduced from the integral equation \rf{cbethe}.
The BMN limit corresponds to a vanishingly small macroscopic density
$\rho(x)$, except for  contributions concentrated
around the zeros of the r.h.s. of \rf{cbethe}:
\be
\frac{1}{{\rm x}_k}=\frac{\kappa}{n_k}
\left(1-\sqrt{1+\frac{n^2_k}{\kappa^2}}\right).
\ee
In terms of the filling fractions \rf{frac},
the normalizations
\rf{nn}, \rf{cbethemom} and \rf{nn1} in this approximation become
\be\label{sums}
\sum_k S_k=\frac{2\pi}{\sqrt{\lambda}}(\Delta+2J-L),
\ \ \ \ \
\sum_k \frac{S_k}{{\rm x}_k}=0,
\ \ \ \ \
\sum_k \frac{S_k}{{\rm x}_k^2}=\frac{2\pi}{\sqrt{\lambda}}(\Delta-L).
\ee
Let us introduce the quantities $N_k$, which parameterize
the filling fractions as
\be
\label{sbmn}
S_k=N_k\frac{2\pi }{\sqrt{\lambda}}
\left(\sqrt{1+\frac{n^2_k}{\kappa^2}}+1\right),
\ee
The $N_k$'s have the meaning of occupation numbers of
string oscillators and should be large for the semiclassical
approximation to work. On the other hand the filling
fractions $S_k$ are small in the approximation we use, so
the coupling must satisfy  $\sqrt{\lambda}\gg N_k$.
The second of the equations \rf{sums} then leads to the
zero-momentum condition in the form
\be
\sum_k N_kn_k=0.
\ee
The total momentum vanishes
in this case, since all $S_k$ are negligibly small, which is consistent
only for $m=0$.

The sum and  difference of the first and third equations of \rf{sums}
gives rise to
\be
J=\sum_k N_k,
\\
\Delta-L=\sum_k N_k\left(
\sqrt{1+\frac{\lambda n^2_k}{\Delta^2}}-1\right),
\ee
which are the familiar BMN results, except that $L$ is replaced by
$\Delta$, but which only leads to corrections
beyond the BMN limit.
{}From the point of view of integrable systems the BMN limit corresponds to
the solitonic degeneration of the finite-gap solutions.
In other words, the cuts shrink to points.
One can also consider partial degenerations
 when some filling fractions are small
but others are arbitrary, which corresponds to
 solitons on the
background of finite-gap solutions. These solutions
describe small oscillations
around macroscopic rotating strings.

\subsection{Rational Solutions
\label{ss:rationals}}

In this subsection we will consider solutions with a single cut.  These
these solutions
are dual to a class of semiclassical string solutions described in
\cite{Arutyunov:2003za}. We will demonstrate that this simple
example is not only instructive for our general approach, but also
show that all rational solutions have two-loop agreement between
the SYM prediction and the semiclassical string prediction.

\paragraph{Spin chain solution  for the one loop gauge theory }\

\noindent
First, we consider the solution to the quasiclassical Bethe equations of
section \ref{ss:generic}.
With  only the one cut ${\bf C}$, eq.~(\ref{BAEC}) reduces to
\be\label{BAEC1}  {1\over x}=2\pi n+
2 \pint_{{\bf C}} {d\xi\rho(\xi)\over x-\xi} =
2\pi n + G(x+i0)+G(x-i0), \ \ \ \ x\in {\bf C}
\ee
The momentum constraint (\ref{MOMG2}) imposes the
further condition that
\be
n\alpha=m
\ee
where $m$ is an integer.  With a single cut, it is straightforward
to find the resolvent $G(x)$ from (\ref{BAEC1}), where one finds
\be\label{Gsol1}
G(x)=\frac{1}{2}\left(\frac{1}{x}-2\pi n+\frac{1}{x}\sqrt{(2\pi nx-1)^2+8\pi
m x}\right)
\ee
i.e. the curve (\ref{sigmaxxx}) in this case is rational, and can be
defined (for $n\neq 0$) by the equation
\be
\label{racu}
y^2 = {1\over 4\pi^2 n^2}\left((2\pi nx-1)^2+8\pi m x\right)=
x^2 - {1\over\pi n}\left(1-{2m\over n}\right)x + {1\over 4\pi^2 n^2}
\ee
The coefficients inside the square root are chosen to eliminate the
pole at $x=0$ and to reproduce the large $x$ behavior of $G$
(\ref{asinf}),
\be
G(x)\stackreb{x\to\infty}{\simeq}\frac{m/n}{x}+\dots
\ee
With the resolvent in hand, we can now find the anomalous dimension,
which is proportional to the linear coefficient of the Taylor
expansion (\ref{gamma}) of $G(x)$ about $x=0$.  Hence,
\be\label{ratgam}
\gamma=\frac{\lambda}{8\pi^2L}\left(-{dG\over dx}\right)_{x=0}
=\frac{\lambda m(n-m)}{2L}
\ee
Comparing this solution to the string solutions in \cite{Arutyunov:2003za},
we see
that this is the same value as the circular string solutions with $J_3=0$
and $J_1$ and $J_2$ satisfying
\be
J_1+J_2=L\\
mJ_1+(m-n)J_2=0
\ee
Notice further that the special case of $n=2$ and $m=1$ is the
original solution of \cite{Frolov:2003qc}.  In other words, the
Frolov-Tseytlin solution may be thought of as a BMN type solution
where there are $\frac{L}{2}$ creation operators
$\alpha^{\dagger}_{+2}$ acting on the chiral primary ground state.
{}From this, it is intuitively clear why this state has a higher energy
than the folded string with the same $R$-charges; the folded string is
made up of $\frac{L}{4}$ creation operators $\alpha^{\dagger}_{+1}$
and $\frac{L}{4}$ creation operators $\alpha^{\dagger}_{-1}$.

The number of Bethe roots is limited to half the number of sites in
the chain.  Therefore, $\alpha=m/n\le\half$.  However, the resolvent
(\ref{Gsol1}) is well behaved if $m>n/2$ and $\gamma$ is invariant
under $m\to n-m$.  This suggests that there is another configuration
of Bethe roots which leads to the same resolvent, but with $\alpha$
for the new configuration given by $1-\alpha$ of the old
configuration.  It is easy to see what the new configuration is by
considering the position of the branch points of (\ref{Gsol1}) which
are located at
\be\label{brpts}
{\rm x}_\pm=\frac{1}{2\pi}\left(\frac{n-2m}{n^2}\pm
\frac{2}{n^2}\sqrt{m^2-nm}\right)
\ee
If $m<n/2$ then the real part in (\ref{brpts}) is positive, but it is
negative if $m>n/2$.  In fact, it is clear that under the transformation
$n\to -n$, $m\to -n+m$, $G(x)$ is transformed as $G(x)\to -G(-x)$ (this
transformation leaves the curve (\ref{racu}) intact).  Hence,
if we start with a configuration with $m>n/2$, this is equivalent to
having $n'=-n$, $m'=-n+m$ and $\alpha=m'/n'<\half$.  In other words, this
corresponds to BMN states with $L(n-m)/n$ creation operators
$\alpha^\dagger_{-n}$ acting on the chiral primary state.

In the case of $n=2m$, the solution  considered here is similar
to the $SO(6)$ singlet solution constructed in \cite{Engquist:2003rn}.
But we could have also chosen the Bethe roots to
consist of a condensate between the branch points and two tails
leading to $\pm i\infty$, this is the solution originally constructed
in \cite{Beisert:2003xu}.  To see how this works,
recall that $\rho(x)dx$ has to be positive
definite along the distribution of the roots.
The special case $n=2m$
has branch points at ${\rm x}_{\pm}=\pm i/(2\pi n)$ and
locally from the branch
points $\rho(x)\sim\sqrt{{\rm x}_\pm-x}$.  Hence there are three directions
coming out of the branch point on which the roots can lie.  Two of these
directions correspond to a cut connecting the two branch points on
either side of the imaginary axis.  The third direction is along
the imaginary axis out to infinity.  This would normally lead to
a nonnormalizable density.  However, we can also choose the opposite
sign of the square root and add a constant density along the imaginary
axis.  This then leads to a condensate between the branch points
and two normalizable tails extending out to infinity.

\paragraph{Solution of the classical  magnet}\

\noindent
The simplest rational solution of the classical ferromagnet equations
in (\ref{krucz}) is
\be\label{ansatz}
\phi=w\tau+n\sigma,\ \ \ \ \ \ \theta=\theta_0,
\ \ \ \ \ {\rm with}
\\
w=n^2\cos\theta_0
\ee
For this solution the total angular momentum is
\be
S_z=\frac{L}{4\pi}\int_0^{2\pi}d\sigma\cos\theta=
\frac{L}{2}\cos\theta_0
\ee
and the anomalous dimension or Hamiltonian is given by
\be
\label{smgamma}
\gamma=\frac{\lambda}{8\pi^2}H=
\frac{\lambda}{16\pi L}\int_0^{2\pi}d\sigma\left[
(\partial_\sigma\theta)^2+\sin^2\theta(\partial_\sigma\phi)^2\right]
=\frac{\lambda n^2}{8L}\sin^2\theta_0
\ee
Now,
for this particular solution one has for $U$ defined in (\ref{UVS})
\be
\left.U\right|_{t=0} =\left(
\begin{array}{cc}
\cos\theta_0 & \sin\theta_0\e^{ in\sigma} \\
 \sin\theta_0\e^{- in\sigma} & -\cos\theta_0
\end{array}
\right)
\ee
and the first of the equations
of the auxiliary linear problem in (\ref{auxili}) can be easily integrated:
\be
\Psi(x;0,\sigma)=\frac{1}{k_+-k_-}\left(
\begin{array}{cc}
k_+\e^{-ip_-\sigma/2\pi}-k_-\e^{-ip_+\sigma/2\pi} &
\frac{\sin\theta_0}{2x}\,\left(\e^{-ip_-\sigma/2\pi}-\e^{-ip_+\sigma/2\pi}\right)
\\
\frac{\sin\theta_0}{2x}\,\left(\e^{ip_+\sigma/2\pi}-\e^{ip_-\sigma/2\pi}\right)
&
k_+\e^{ip_-\sigma/2\pi}-k_-\e^{ip_+\sigma/2\pi}
\end{array}
\right),
\ee
where
\be\label{quasir}
k_\pm= \pi n\left(\frac{\cos\theta_0}{2\pi nx}-1\pm\frac{1}{x}\,
\sqrt{x^2-{\cos\theta_0\over\pi n}\,x+{1\over 4\pi^2n^2}}\right)\,,
\\
p_\pm=\pi n\left(\pm\frac{1}{x}\,
\sqrt{x^2-{\cos\theta_0\over\pi n}\,x+{1\over 4\pi^2n^2}}-1\right).
\ee
are functions on (\ref{racu}).
Putting $\sigma=2\pi$, taking the trace and observing that $\cos  p_+=\cos
p_-$, we get
\be
{\rm Tr}\,\Psi(x;0,2\pi)={\rm Tr}\,\Omega=2\cos  p_\pm
\ee
Hence, the quasi-momentum on the physical sheet
is $p_+$ in (\ref{quasir}).

We now have to impose the momentum constraint.  One way to find
the value of the
momentum is to compute the expectation value of a shift operator for a quantum
state.  The correspondance principle then relates this to the classical value.
The quantum state is a collection of $L$ ``up'' spins, where ``up'' for the
$j^{\rm th}$ spin is
with respect to a polar and an azimuthal angle $\theta_j$ and $\phi_j$.  Hence
the state has the form
\be
|{\rm state}\rangle=\left|\left(\begin{array}{c}
\cos{\theta_1\over 2}\\e^{i\phi_1}\sin{\theta_1\over 2}
\end{array}\right)...\left(\begin{array}{c}
\cos{\theta_j\over 2}\\e^{i\phi_j}\sin{\theta_j\over 2}
\end{array}\right)...\left(\begin{array}{c}
\cos{\theta_L\over 2}\\e^{i\phi_L}\sin{\theta_L\over 2}
\end{array}\right)\right\rangle.
\ee
and it is parameterized, as usual, by spinor ``half-angles"
$\theta/2$ which in
our case correspond to the sigma-model co-ordinates given below.

The shift operator $e^{iP}$
sends the spin at site $j$ to the spin at site $j+1$.
Hence, the expectation value of the shift operator is
\be
\langle {\rm state}|e^{iP}|{\rm state}\rangle=
\prod_{j=1}^L\left(\cos\frac{\theta_j}{2}\cos\frac{\theta_{j+1}}{2}+
\sin\frac{\theta_j}{2}\sin\frac{\theta_{j+1}}{2}e^{i(\phi_{j+1}-\phi_j)}
\right)\approx
\\
\approx\prod_{j=1}^L\left(1+i\sin^2{\theta_j\over 2}\Delta\phi_j\right)
\approx \exp\left(i\sum_j\sin^2\frac{\theta_j}{2}\Delta\phi_j\right),
\ee
where we have assumed that $\Delta\theta_j<<1$, $\Delta\phi_j<<1$, which
is valid in the long wavelength limit.
Hence, the classical momentum is given by
\be\label{mom}
P=\int_0^{2\pi}d\sigma \sin^2\frac{\theta}{2}\partial_\sigma\phi=
2\pi n\sin^2\frac{\theta_0}{2}.
\ee
The momentum is required to be a multiple of $2\pi$, which puts the restriction
on $\theta_0$,
\be
\cos\theta_0=1-{2m\over n}
\ee
showing that $p_+$ in (\ref{quasir}) is consistent with $G(x)$ in
(\ref{Gsol1}).

Let us now write everything in terms of the $R$-charges $J_1$ and $J_2$.
We have that
\be
J_1={L\over 2}+S_3,\qquad
J_2={L\over 2}-S_3=\frac{L}{2}(1-\cos\theta_0)=L\sin^2{\theta_0\over 2}.
\ee
We also have that $\alpha=J_2/L$, hence, $P$ is given by
\be
P=2\pi n \alpha,
\ee
and the momentum condition then forces $\alpha n=m$ where $m$ is an integer.
This then leads to
\be
\gamma=\frac{\lambda m(n-m)}{2L}
\ee
which is the same as (\ref{ratgam}).
Note that the singly wound Frolov-Tseytlin
circular string reduces to a doubly wound classical
solution of the XXX ferromagnet.

\paragraph{Solution of the sigma-model}\

\noindent
The string
is restricted to live on $S^3$, so we choose the parameterization
\be
Z_1=X_1+iX_2,\qquad\qquad Z_2=X_3+i X_4,
\ee
where
\be\label{unity}
|Z_1|^2+|Z_2|^2=1.
\ee
Let us consider solutions of the form
\be\label{stringanstz}
Z_1=\cos{\theta_0\over 2} e^{iw_1\tau+i m_1\sigma},\qquad\qquad
Z_2=\sin{\theta_0\over 2} e^{iw_2\tau+i m_2\sigma},\qquad t=\kappa\tau.
\ee
which, in particular, corresponds to the rational case
\rf{psira} of the general construction of Appendix~\ref{ss:kri}.
The angle $\theta_0/2$ is
the polar angle on $S_3$ in generalized coordinates and
so $\theta_0$ ranges between $0$ and $\pi$.
The equations of motion for $Z_{1,2}$ lead to the relation
for the parameters of the solution
\be
\label{restr}
w_1^2-m_1^2=w_2^2-m_2^2,
\ee
while the Virasoro constraints (\ref{virasoro}) require
\be
\label{vir1}
\kappa^2=(w_1^2+m_1^2)\cos^2{\theta_0\over 2}+
(w_2^2+m_2^2)\sin^2{\theta_0\over 2}
\\
w_1m_1\cos^2{\theta_0\over 2}+w_2m_2\sin^2{\theta_0\over 2}=0.
\ee
in particular meaning that
\be
\label{cosi}
\cos^2{\theta_0\over 2} = {w_2m_2\over w_2m_2-w_1m_1},
\ \ \ \ \
\sin^2{\theta_0\over 2} = {w_1m_1\over w_1m_1-w_2m_2}
\ee
and
\be
\kappa^2 = {(w_1w_2+m_1m_2)(w_1m_2-w_2m_1)\over w_2m_2-w_1m_1}
\ee
which is, together with (\ref{restr}) solved by
\be
\label{wmW}
w_I = {\kappa\over 2}\left(\sqrt{W_I}+{1\over\sqrt{W_I}}\right)
\\
m_I = {\kappa\over 2}\left({1\over\sqrt{W_I}}-\sqrt{W_I}\right)
\ee
which is equivalent to \rf{kraW}. Using  \rf{wmW}, the formulae in \rf{cosi}
become equivalent to \rf{rrat}.

The energy, which is $\Delta$, and the $R$-charges of the string are given by
\be
\Delta=\sqrt{\lambda}\kappa\nonumber\\
J_1=\sqrt{\lambda}\int_0^{2\pi}\frac{d\sigma}{2\pi}\cos^2{\theta_0\over 2}w_1=
\sqrt{\lambda}\cos^2{\theta_0\over 2}w_1
\\
J_2=\sqrt{\lambda}\int_0^{2\pi}\frac{d\sigma}{2\pi}\sin^2{\theta_0\over 2}w_2=
\sqrt{\lambda}\sin^2{\theta_0\over 2}w_2
\ee
Now, the relations (\ref{restr}) and (\ref{vir1}) lead to the equations
\be
(w_1^2+m_2^2-m_1^2)\left(J_1 -\sqrt{\lambda}w_1\right)^2-
\left(J_2w_1\right)^2=0
\\
J_1m_1+J_2m_2=0
\ee
Note that the situation simplifies greatly if $m_1=-m_2=m$, in which case
we find that
\be\label{simplerel}
\Delta^2=(J_1+J_2)^2+\lambda m^2.
\ee
In order to compare this to a single cut solution of the Bethe
equations, it is convenient
to rescale $x$ by a factor of $a={1\over 4\pi \kappa}$.  Equations
(\ref{compa}) are then modified to
\be
\label{compa1}
\int dx\,\rho (x)=
\frac{J}{\Delta}+\frac{\Delta-L}{2\Delta }\,,
\\
-\oint {dx\over 2\pi ix}\,G(x) = 2\pi m,
\\
- \frac{\lambda }{8\pi ^2\Delta}\oint {dx\over 2\pi ix^2}\,G(x)=\Delta -L,
\ee
and (\ref{cbethe}) becomes
\be\label{cbetherat}
G(x+i0)+G(x-i0) =
2\pint d\xi\,\frac{\rho(\xi)}{x-\xi}=\frac{1}{2}\left(\frac{1}{x-a}
+\frac{1}{x+a}\right)-2\pi n,
\ee
Then the general form of $G(x)$ for
$m_1=m$, $m_2=m-n$, satisfying \rf{cbetherat} and
compatible with the general solution \rf{twopole}
for the chiral field, is
\be\label{Geq}
G(x)=\frac{1}{4}\left(\frac{1}{x-a}+\frac{1}{x+a}\right)
+\frac{1}{4}\left(\frac{(1+\eps)^{-1/2}}{x-a}+
\frac{(1-\eps)^{-1/2}}{x+a}\right)\sqrt{Ax^2+Bx+C}-\pi n,
\ee
where, in order to cancel the poles at $x=\pm a$ on the physical sheet,
we must satisfy the relations
\be\label{bcarel}
B=4\pi\kappa\eps\\
C+\frac{A}{(4\pi\kappa)^2}=1.
\ee
In order to have the  asymptotic behavior of (\ref{asinf})
we also require
\be\label{arel}
\sqrt{A}\left(\frac{1}{\sqrt{1+\eps}}+\frac{1}{\sqrt{1-\eps}}\right)=4\pi n,
\ee
and to satisfy the momentum condition (\ref{cbethemom}) we have
\be\label{crel}
G(0)=\pi\kappa\sqrt{C}\left(\frac{1}{\sqrt{1+\eps}}-
\frac{1}{\sqrt{1-\eps}}\right)
-\pi n=-2\pi m.
\ee
If we define $Y\equiv\sqrt{1-\eps^2}$, then equations
(\ref{bcarel}), (\ref{arel}) and (\ref{crel}) lead to the equation
\be\label{kapparel}
\kappa^2=\frac{Y^2}{2}\left(\frac{n^2}{1+Y}+\frac{(n-2m)^2}{1-Y}\right).
\ee
Using (\ref{bcarel}), (\ref{arel}) and the definition for
$Y$, one can show that asymptotically, $G(x)$ in (\ref{Geq}) behaves as
\be\label{Gasymp}
G(x)\ \stackreb{ x\to\infty}{\sim}\
\frac{1}{2x}+\left(\frac{\kappa^2(1+Y)\sqrt{1-Y^2}}{Y^2}-
\frac{n^2(1-Y)}{\sqrt{1-Y^2}}\right)\frac{1}{4n\kappa x}.
\ee
Substituting (\ref{kapparel}) for $\kappa^2$, and comparing
(\ref{Gasymp}) with  (\ref{compa1})
leads to the equation
\be\label{ynorm}
n^2(1-Y)^2-(n-2m)^2(1+Y)^2=\frac{4n(J_1-J_2)}{\sqrt{\lambda}}\sqrt{1-Y^2}.
\ee
It is then a straightforward excercise to show that (\ref{compa1})
is consistent with (\ref{kapparel}) and (\ref{ynorm}) if
\be
J_1=\frac{n-m}{n}L\qquad\qquad J_2=\frac{m}{n}L.
\ee
Hence, we have that $m=m_1$,
$n=m_1+m_2$ and $Y$ satisfies the quartic equation
\be\label{yquart}
\left(n^2(1-Y)^2-(n-2m)^2(1+Y)^2\right)^2-
16(n-2m)^2(1-Y^2)\frac{L^2}{\lambda}=0.
\ee
In the special case where $n=2m$, then $Y=1$ and (\ref{kapparel}) reduces to
(\ref{simplerel}).  We also see using (\ref{simplerel})
that $G(x)$ in (\ref{Geq})
simplifies to
\be\label{Gsimp}
G(x)=\frac{1}{2}\frac{x}{x^2-{\lambda\over 16\pi^2}{1\over L^2+\lambda m^2}}
\left(1+ 4\pi m
\sqrt{x^2+{1\over 16\pi^2m^2}\frac{L^2}{L^2+\lambda m^2}}\,\right)-2\pi m.
\ee
Finally, expanding $G(x)$ about $x=0$ and matching the linear term to
the negative of the third equation in
(\ref{compa1}), one immediately finds (\ref{simplerel}).

\paragraph{Solution of the spin chain at two loops and comparison with the sigma-model}\

\noindent
For general values of $n$ and $m$, we can at least solve for $\Delta$
in a series expansion of $\lambda/L^2$.  Solving for $Y$ in
(\ref{yquart}) we find
\be
Y=1-\frac{(n-2m)^2}{2}\frac{\lambda}{L^2}+
\frac{3(n-2m)^4}{8}\frac{\lambda^2}{L^4}-
\frac{(6(n-2m)^2-n^2)(n-2m)^4}{16}\frac{\lambda^3}{L^6}+...
\ee
Substituting this into (\ref{kapparel}) we find
\be\label{2looprat}
\Delta=L+\frac{m(n-m)}{2}\frac{\lambda}{L}-\frac{m(n-m)(n^2-3m(n-m))}{8}\frac{\lambda^2}{L^3}+...
\ee
The term linear in $\gamma$ in (\ref{2looprat}) is clearly consistent with
(\ref{ratgam}) but now let us consider the two loop term.  To do this,
we consider the one cut solutions to the two loop approximation
of the Inozemtsev chain.  Using (\ref{BAETL}), we can write down a general
solution for $G(x)$ compatible with \rf{DPCORR} as
\be\label{GratIn}
G(x)=\frac{1}{2x}+\frac{T}{x^3}+
\left(\frac{D}{2x}+\frac{E}{x^2}+\frac{T}{x^3}\right)
\sqrt{Ax^2+Bx+1}-\pi n,
\ee
where we recall that $T={\lambda\over 16\pi^2L^2}$.
Cancellation of the double and single pole at $x=0$ leads to the equations
\be
E+\frac{1}{2}BT=0,
\ee
\be
D+EB+AT-\frac{1}{4}B^2T=1.
\ee
The momentum condition (\ref{QMQCL}) to linear order in $\lambda/L^2$
gives
\be
-\frac{1}{4}DB+\frac{1}{8}EB^2-\frac{1}{2}EA-\frac{1}{16}TB^3+\frac{1}{4}TAB
-T\left(\frac{1}{16}B^3-\frac{1}{4}AB\right)
=\pi(n-2m),
\ee
where the parentheses term comes from the second term in
(\ref{QMQCL}).
Finally, the constraint on the constant term at infinity is
\be
D\sqrt{A}=2\pi n.
\ee
Solving these equations up to terms linear in $\lambda/L^2$, we find
\be
D=1-\delta, \ \ \ \ E=\frac{\lambda}{4\pi L^2}(n-2m), \ \ \ \
A=(2\pi n)^2(1+2\delta),\\
B=-4\pi(n-2m)\left(1+\delta+4m(n-m)\frac{\lambda}{L^2}\right),
\ee
where
\be
\delta=-\frac{\lambda}{2L^2}(n^2-6m(n-m)).
\ee
Inserting these values into (\ref{GratIn}), we find that
\be\label{Gp}
G'(0)=-4\pi^2m(n-m)+2\pi^2m(n-m)(n-3m)(2n-3m)\frac{\lambda}{L^2}+...
\\
\frac{1}{3!}G'''(0)= -16\pi^4m(n-m)(n^2-5m(n-m))+...
\ee
Applying this to (\ref{GAMCOR}), we finally obtain
\be\label{gInrat}
\gamma=-\frac{\lambda}{8\pi^2L}G'(0)-\frac{3\lambda^2}{128\pi^4L^3}
{G'''(0)\over 3!}+\dots=\frac{m(n-m)}{2}\frac{\lambda}{L}
-\frac{m(n-m)(n^2-3m(n-m))}{8}\frac{\lambda^2}{L^3}+\dots,
\ee
which, finally, agrees with (\ref{2looprat}).

\subsection{Pulsating Solutions}

This section is somewhat outside the main development of this paper,
but it indicates that similar techniques may be applied to operators
that are not of the XXX type.
The relevant observation is that the isometry group of the sigma model
on $S^3$ is $SU(2)_R\times SU(2)_L$, while the isometry group of the
Heisenberg magnet is $SU(2)_R\times U(1)_L$.  Hence, there will be
classical solutions of the sigma model that do not reduce to
solutions of the Heisenberg magnet as the coupling is taken to zero.
The solutions that do reduce to the magnet are those where $Q_L$ equals
the bare dimension of the operator.

We now relax this condition and
consider the so called pulsating  string solutions.
These string solutions are not dual to XXX operators, but instead
are dual to operators comprised of all six types of scalar fields.
To proceed we consider
\be\label{pulseansatz}
Z_1=\cos\chi e^{im\sigma},\qquad\qquad Z_2=\sin\chi e^{i\phi},
\ee
where $\chi$ and $\phi$ are functions only of $\tau$.
The Virasoro constraints \rf{virasoro} give
\be\label{virpulse}
\kappa^2=\dot\chi^2+m^2\cos^2\chi+\dot\phi^2\sin^2\chi.
\ee
We also have from the equations of motion that
\be
-\ddot\chi\tan\chi-\dot\chi^2+m^2=
\ddot\chi\cot\chi-\dot\chi^2-\dot\phi^2+2i\dot\chi\dot\phi\cot\chi
+i\ddot\phi.
\ee
Setting the imaginary piece to zero and integrating leads to the equation
\be
\dot\phi=\frac{J/\sqrt{\lambda}}{\sin^2\chi},
\ee
where the integration constant is determined by the $R$-charge $J$.
If we now insert (\ref{pulseansatz}) into (\ref{su2}) and (\ref{laxcf}) we
find the auxiliary linear equation
\be
\partial_\sigma\Psi(x;0,\sigma)=\frac{x}{x^2-1}\left(
\begin{array}{cc}
-iJ/\sqrt{\lambda} & e^{-im\sigma}\dot\chi\\
-e^{im\sigma}\dot\chi& iJ/\sqrt{\lambda}
\end{array}\right) \Psi(x;0,\sigma),
\ee
where we have assumed that $\left.\phi\right|_{\tau=0}=
\left.\cos\chi\right|_{\tau=0}=0$.  Integrating this
equation and using the Virasoro constraint in (\ref{virpulse})
we find that the quasi-momentum is given by
\be
p(x)=\frac{2\pi}{x^2-1}\sqrt{\left(\frac{m}{2}(x^2-1)-
{J\over\sqrt{\lambda}}x\right)^2+
\left(\kappa^2-{J^2\over\lambda}\right)x^2} -\pi m\,.
\ee
and the resolvent $G(x)$ is
\be
G(x)=\frac{2\pi}{x^2-1}\left(\kappa
x+\sqrt{\left(\frac{m}{2}(x^2-1)-{J\over\sqrt{\lambda}}x\right)^2+
\left(\kappa^2-{J^2\over\lambda}\right)x^2}\,\right) -\pi m\,.
\ee
Rescaling again $x$ by $4\pi\kappa$, we get for $G(x)$
\be\label{pulseallloop}
G(x)=\frac{1}{2}\frac{1}{x^2-{\lambda\over 16\pi^2\Delta^2}}
\left( x+\sqrt{\left(2\pi m
\left(x^2-{\lambda\over 16\pi^2\Delta^2}\right)-(1-\beta) x\right)^2+
\beta(2-\beta)x^2}\,\right) -\pi m\,.
\ee
where $\beta=1-{J\over\Delta}$.  When $\lambda\to 0$ this becomes
\be\label{pulse1loop}
G_0(x)=\frac{1}{2x}\left( 1+\sqrt{\left(2\pi mx-(1-\beta) \right)^2+
\beta(2-\beta)}\,\right) -\pi m\, .
\ee
In \cite{Engquist:2003rn}, it was shown that the resolvent coming from the
one-loop Bethe equations for $[0,J,0]$ representations of $SO(6)$
was proportional to $G_0(x)-G_0(-x)$.
In terms of (\ref{pulse1loop}) we
see that this corresponds to adding together the resolvents for two
classical solutions with $\pm m$.  This suggests that the full resolvent,
dual to these states, is $G(x)-G(-x)$ where $G(x)$ is given by
(\ref{pulseallloop}). It is also interesting to point out, that at the
limit $\lambda\to 0$ the elliptic curve of \rf{pulseallloop} turns into
the rational of \rf{pulse1loop}, which is reminiscent
of a corresponding phenomenon
in softly broken \4N Seiberg-Witten theory.

\subsection{Elliptic Solutions
\label{ss:ellip}}

The next case in simplicity after the rational solutions of Bethe equations
are solutions with roots distributed on two symmetric cuts.
The Riemann surface (\ref{sigmaxxx}) of a two-cut solution
\be
\label{elcu}
y^2=(x^2-{\rm x}_1^2)(x^2-{\rm x}_2^2)
\ee
is a torus and is just a particular $K=2$ case of (\ref{sigsym}).
The resolvent for the two-cut solutions can be expressed
in terms of elliptic integrals. Elliptic solutions are parameterized by
a pair of integers, $n$ and $m$, which determine two independent
periods of the differential $dp$. Two types of elliptic solutions
found in~\cite{Beisert:2003xu} are dual to circular and
to folded strings rotating in $S^5$. We will show these solutions
 essentially exhaust all possiblities.
In the process we will show that not all pairs of integers
$n$ and $m$ are compatible with the reality condition and that some pairs
are related by symmetries and correspond to one and the same distribution
of Bethe roots.

According to sect.~\ref{ss:generic},
two-cut solutions of Bethe equations are constructed with the help
of a meromorphic differential on a double cover of the complex plane
with two cuts $({\rm x}_1,{\rm x}_2)$ and $(-{\rm x}_2,-{\rm x}_1)$.
The differential with the desired properties is
\be\label{diffal}
dp=-\frac{dx\left(\frac{1}{2}-\alpha
-\frac{{\rm x}_1{\rm x}_2}{2x^2}\right)}
{\sqrt{({\rm x}_2^2-x^2)(x^2-{\rm x}_1^2)}}
\ee
The constant term in the numerator assures the correct normalization at
infinity
 (\ref{infty}). The square root is defined in the standard way,
such that it coincides with its algebraic value at infinity. The value of
$\sqrt{(x^2-{\rm x}_1^2)(x^2-{\rm x}_2^2)}$ at $x=0$ depends on how the complex plane is cut.
We assume that $\sqrt{(0^2-{\rm x}_1^2)(0^2-{\rm x}_2^2)}=-{\rm x}_1{\rm x}_2$,
which is the standard convention for real positive ${\rm x}_1$ and ${\rm x}_2$.
The anomalous dimension (\ref{gamma})
in this case is
\be
\label{adimel}
\gamma =\frac{\lambda}{16\pi ^2L}
\left(\frac{1-2\alpha }{{\rm x}_1{\rm x}_2}-\frac{{\rm x}_1^2+{\rm x}_2^2}{2{\rm x}_1^2{\rm x}_2^2}\right)
\ee
The periods of $dp$ can be expressed through the complete
elliptic integrals of the first and second kind.
In order to convert them into the standard Legendre form, we first
compute the integrals for positive real ${\rm x}_1$ and ${\rm x}_2$  and then analytically
continue
to complex values of the parameters:
\be
\oint_{A} dp = -2i\int_{{\rm x}_1}^{{\rm x}_2}dx\,\frac{\frac{1}{2}-\alpha
-\frac{{\rm x}_1{\rm x}_2}{2x^2}}
{\sqrt{({\rm x}_2^2-x^2)(x^2-{\rm x}_1^2)}}
=-\frac{i}{{\rm x}_2}\left[(1-2\alpha )K(1-r)-\frac{E(1-r)}{\sqrt{r}}\right]
\\
\oint_{B} dp = 4\int_{{\rm x}_2}^{\infty}dx\,\frac{\frac{1}{2}-\alpha
-\frac{{\rm x}_1{\rm x}_2}{2x^2} }{\sqrt{(x^2-{\rm x}_1^2)(x^2-{\rm x}_2^2)}}
=\frac{2}{{\rm x}_2}\left[(1-2\alpha )K(r)+\frac{E(r)-K(r)}{\sqrt{r}}\right]
\ee
where
the modulus of elliptic integrals is
\be
r=\frac{{\rm x}_1^2}{{\rm x}_2^2}
\ee
Our conventions about elliptic integrals and some of their useful
properties are listed
in the Appendix~\ref{ss:ellint}.

If we assign the momentum quantum numbers $n$ and $-n$ to the two cuts,
so that the B-period of $dG$ is normalized
to $4\pi n$, and put a condensate between the cuts,
such that the A-period  is  equal to $2\pi m$.
The conditions \rf{ABint} then become
\be\label{intcnd}
(1-2\alpha )K(1-r)-\frac{E(1-r)}{\sqrt{r}}=2\pi i m{\rm x}_2
\\
(1-2\alpha )K(r)+\frac{E(r)-K(r)}{\sqrt{r}}=2\pi n{\rm x}_2
\ee
Exluding ${\rm x}_2$ from these equations we find
\be\label{1-2alpha}
1-2\alpha=\frac{1}{\sqrt{r}}\,\frac{nE(1-r)+im\left(E(r)-K(r)\right)}
{nK(1-r)-imK(r)}
\ee
This equation determines the modular parameter $r$ as a function
of the filling fraction $\alpha$. Once $r$ is known, we can
express the end-points of the cuts, and therefore the resolvent
and the anomalous dimension, as implicit functions of $\alpha$.
Using the Legendre relation (\ref{Legendre}) in intermediate
calculations, we find
\be\label{ab}
{\rm x}_1=\frac{1}{4}\,\frac{1}{nK(1-r)-imK(r)}
\\
{\rm x}_2=\frac{1}{4\sqrt{r}}\,\frac{1}{nK(1-r)-imK(r)}
\ee
The above equations in principle solve the problem of computing
the anomalous dimension
as a function of the filling fraction. However, the anomalous dimension
must be real which is not always true for arbitrary $n$ and $m$.
The reality condition constraints allowed values of $n$ and $m$.
There are also certain symmetries
that relate solutions with different $n$ and $m$.

The modular parameter $r$ is an auxiliary variable in the above equations,
it just parameterizes the dependence of ${\rm x}_1$
and ${\rm x}_2$ on $\alpha$. Nothing changes if $r$ is replaced with $1/r$.
The elliptic integrals tansform linearly under such transformation,
according to
(\ref{Landen}). As a consequence,  (\ref{1-2alpha}), (\ref{ab}) have almost
an identical form when expressed in terms of $1/r$.
A short calculation shows that the replacement of $r$
by $1/r$ results in a simple transformations of $n$, $m$, ${\rm x}_1$ and
${\rm x}_2$:
\be\label{symm1}
n\rightarrow n-m,\nonumber \\
{\rm x}_1\rightarrow {\rm x}_2, \nonumber \\
{\rm x}_2\rightarrow {\rm x}_1.
\ee
The resolvent and the anomalous dimension
are manifestly symmetric under interchange of ${\rm x}_1$ and ${\rm x}_2$. Therefore
two pairs of integers, $(n,m)$ and $(n-m,m)$ correspond to one
and the same solution of the Bethe equations.
Hence, $n$ is
only defined modulo $m$.

Complete elliptic integrals $K(r)$ and $E(r)$ are multivalued functions
of $r$. They have
 a logarithmic branch point at $r=1$. We can pick a different branch
of $K(1-r)$ and $E(1-r)$
by pulling $r$ around zero.
$K(r)$ and $E(r)$ are analytic at zero and do not change,
but $K(1-r)$
and $E(1-r)$ pick up discontinuities at the logarithmic cut. The
discontinuities can be easily calculated from
the defining integrals (\ref{ellint}):
$$K(1-r-i0)-K(1-r+i0)=-2iK(r)$$ and
$$E(1-r-i0)-E(1-r+i0)=2i\left(E(r)-K(r)\right).$$
As a result,
\be
K(1-r)\rightarrow K(1-r)-2iK(r),  \\
E(1-r)\rightarrow E(1-r)+2i\left(E(r)-K(r)\right).
\ee
This transformation, when applyed to (\ref{1-2alpha}), (\ref{ab}),
shifts $m$ by $2n$. This is not the only effect of the analytic continuation.
The $\sqrt{r}$ also changes sign when $r$ encircles zero. Therefore
${\rm x}_2$ and $1-2\alpha$ will change sign:
\be\label{trans}
m\rightarrow m+2n, \\
{\rm x}_1\rightarrow {\rm x}_1, \\
{\rm x}_2\rightarrow -{\rm x}_2, \\
\alpha\rightarrow 1-\alpha.
\ee
To get the physical filling fraction ($\alpha<\half$) after the
transformation,
the initial solution
should have $\alpha>\half$. It thus corresponds
to an analytic continuation of some eligible solution beyond $\alpha=\half$.
The anomalous dimension (\ref{adimel}) is manifestly invariant under
(\ref{trans}).
The resolvent is also invariant because the change of sign of the
numerator in \rf{diffal} is compensated by switching to another branch
of the square root in the denominator. Indeed,
the cuts connect ${\rm x}_1$ with $-{\rm x}_2$ and ${\rm x}_2$ with $-{\rm x}_1$ after the transformation,
rather than ${\rm x}_1$ with ${\rm x}_2$ and $-{\rm x}_2$ with $-{\rm x}_1$. Such a rearrangement of the cuts
changes the sign of the square
root $\sqrt{({\rm x}_2^2-x^2)(x^2-{\rm x}_1^2)}$.
The symmetries \rf{symm1} and \rf{trans}
correspond to transformations
in the homology basis:
$(A,B)\rightarrow(A,B-2A)$ and $(A,B)\rightarrow(A+B,B)$.
While the former transformation is a genuine symmetry,
the latter changes the filling fraction and maps a physical solution
of Bethe equations to an analytic continuation of that or of some other
solution.

Apart from integrality of periods of $dp$,  solutions of Bethe
equations must satisfy
the reality condition on ${\rm x}_1$ and ${\rm x}_2$. There are four possibilities: (i)
${\rm x}_1=\overline{{\rm x}_2}$,
 (ii) ${\rm x}_1$, ${\rm x}_2$ are pure imaginary,
(iii) ${\rm x}_1=-\overline{{\rm x}_2}$, and (iv) ${\rm x}_1$, ${\rm x}_2$ are real.
 There are no solutions with real cuts,
which excludes case (iv), but it is known that such solutions
exist for spin $s=-\half$ Heisenberg chain that is also
relevant for the AdS/CFT correspondence  \cite{Beisert:2003ea}
and describes strings rotating in $S^5$ and $AdS_5$
simultaneously  \cite{Frolov:2002av}.
The other three cases are possible for various $n$ and $m$.
Careful inspection of eqs. (\ref{1-2alpha}), (\ref{ab}) shows that there are
three series of solutions that satisfy the reality condition (see
Appendix~\ref{ss:real}
for details): (i) $m=0$, (ii)  $n=0$
and (iii) $m=2n$. The first two cases were studied in \cite{Beisert:2003xu},
and it is straightforward to check that the general formulae here reduce to
those in \cite{Beisert:2003xu} when either $m$ or $n$ is set to
zero\footnote{It is easier to compare to \cite{Arutyunov:2003rg}, where
all elliptic integrals are reduced to the standard Legendre form.}. We should
mention that only the
minimal case of $m=2$ has been discussed in
\cite{Beisert:2003xu,Arutyunov:2003rg}.
The reason why Bethe state with
$m=1$, which has lower energy, is not minimal is that it violates the momentum
constraint,
since the total momentum of Bethe states in case (ii) is $\pi m$.

The case (iii) can be obtained by applying tranformation (\ref{trans})
to case (i) which should be analytically continued past $\alpha=\half$,
because
the transformation maps $\alpha$ to $1-\alpha$. It was conjectured in
 \cite{Beisert:2003ea} that such an analytic continuation
describes  an ``umbrella''-like distribution of Bethe roots, with the
condensate
along the imaginary axis and two  arrays of roots at both ends of the
condensate
which are approximately parallel to the real
axis  (that is, with ${\rm x}_1=-\overline{{\rm x}_2}$). Such states
 were found numerically for small lattices
\cite{Beisert:2003xu}. The solution with $m=2n$ has all the expected
properties:
it is equivalent to an analytic continuation of the solution with $m=0$,
it contains a condensate, and the tails of the
condensate have end-points ${\rm x}_1=-\overline{{\rm x}_2}$.
We thus conclude that the elliptic solution of Bethe equations with
$m=2n$ is the ``umbrella" state conjectured in  \cite{Beisert:2003ea}.
It was also observed in  \cite{Beisert:2003xu,Beisert:2003ea}
that the solution with $m=0$
(case (ii)) is symmetric under $\alpha\rightarrow 1-\alpha$.
{}From the point of view of the general approach developed here,
this is a consequence of the
invariance under the transformation
\rf{trans}.



\sectiono{Discussion
\label{ss:discussion}}

In this paper we have discussed a general, integrability
based  approach for connecting the
two sides of the gauge-string duality. On
the gauge side we have composite operators, whose renormalization is
governed by a quantum integrable spin chain, while on the string side we have
a classically integrable  sigma-model.

Our explicit calculations were mostly
restricted to the $SU(2)$ subsectors of  SYM and string
theory. The one- and two-loop operator mixing in the $SU(2)$  sector can be resolved for
arbitrarily large
operators and expressed in terms of simple geometric data. A particular
operator is characterized by a hyperelliptic complex
curve, endowed with extra meromorphic differential, whose periods obey
certain integrality conditions.
It turns out that solutions of the sigma-model
are expressed in terms of the same analytic data, with a slight difference
in the pole structure
of the differential. This difference however accounts for all orders
in the BMN coupling $\lambda/L^2$. On the gauge theory side, the geometric
data is derived by solving the scaling limit of the
discrete Bethe equations, which
are themselves exact, and not restricted to operators of
large size.
The discrete counterpart of the geometric data in the sigma model,
if it exists,
would describe the full non-perturbative spectrum of single-trace operators in
the SYM theory.
Presumably, it can be derived from quantum  string theory in
$AdS_5\times S^5$, about which little is known. Our results suggest that
the quantum
sigma-model can likely be  solved using a  Bethe ansatz
whose scaling limit produces the
integral equation \rf{cbethe}. As sometimes happens for integrable systems,
 the full quantum solution of the model may even be
uniquely determined by its semiclassical limit and an underlying analytic
structure.

The energy of the string solutions is analytic in the 't~Hooft
coupling $\lambda$
and can be expanded in a power series in $\lambda/L^2$. This expansion,
though being very nontrivial on the string side,
agrees with SYM perturbation theory up to two loops. We proved the agreement
for the general solution in the $SU(2)$ sector.
The two-loop agreement was also observed (for BMN states)
at the level of $1/L$ corrections  \cite{Beisert:2003tq,Callan:2003xr},
which are described by
quantum corrections in the sigma-model.
In view of the discrepancies found in \cite{Serban:2004jf} for elliptic solutions
at three loops, these results should be taken cautiously. Definitely more
work is needed to understand if string solitons produce all-order non-perturbative
results or the agreement is restricted to the first few orders of perturbation theory.
We hope that the approach based on
integrability is the right framework to address this question.

 It would be
also interesting to generalize the approach developed in this paper to all scalar
operators
\cite{Minahan:2002ve}
 or even to the full set of single-trace operators in the SYM theory
\cite{Beisert:2003yb}, for which the one-loop Bethe equations are known.
It is natural to guess that the solution should be expressed
in terms of differentials on more complicated, non-hyperelliptic Riemann
surfaces. Again, we
expect more complicated analytic structures to arise there, in agreement
with the integrable structure of the full $AdS_5\times S^5$ sigma models
which were only briefly discussed in this paper.
We hope to return to these
problems elsewhere.

\paragraph{Note added:} 
After we originally submitted this paper, we received \cite{GS} where the transformation
that maps B\"acklund charges in the sigma-model to those in the spin system
was extended to two loops.  We later received \cite{Kruczenski:2004kw}, where
it was shown that the effective action for the Heisenberg chain can
be mapped onto the string action to at least two-loop order. 


\subsection*{Acknowledgements}

We would like to thank N.~Beisert, L.~Faddeev, A.~Gorsky, I.~Gruzberg,
G.~Korchemsky, I.~Kostov, I.~Krichever, D.~Kutasov, S.~Lukyanov,
N.~Nekrasov, A.~Parnachev, A.~Ryzhov, N.~Slavnov, F.~Smirnov, M.~Staudacher,
A.~Tseytlin and P.~Wiegmann for discussions. V.K. is  indebted
to F.Smirnov for explaining to him the papers \cite{Smirnov:1998kv} and
\cite{RESHSM}.  A.M. is grateful to A.~Niemi
and the Department of Theoretical Physics at Uppsala University
for the kind hospitality , where
this work started, and to IHES, where it has been completed.
K.Z. would like to thank NHETC at Rutgers,
where part of this work was done, for warm hospitality. The work of V.K. was partially
supported by European Union under the RTN contracts HPRN-CT-2000-00122
and 00131, and by NATO grant PST.CLG.978817. The work of A.M. was
partially supported by STINT grant IG2001-062, the RFBR grant
03-02-17373, the program of support of scientific schools 1578.2003.2,
and by the Russian science support foundation. J.A.M. and K.Z. were
supported in part by the Swedish Research Council.  J.A.M. was also
supported in part by DOE contract \#DE-FC02-94ER40818.

\appendix

\sectiono{General finite-gap solutions of the sigma-models
\label{ss:kri}}

An alternative and more generic way to construct the
finite-gap solutions for relativistic two-dimensional
sigma-models, satisfying the Virasoro constraints,
was proposed by Krichever in \cite{Krisig}. Here we review this approach
and modify it in order to satisfy the Virasoro constraints \rf{virasoro}
(with the non-vanishing r.h.s.). Moreover, some of the formulas of this
construction appear to be useful for the examples we consider in
sect.~\ref{ss:rationals}, and should be even more useful for the
generalizations of our results beyond the $SU(2)$ sector.

The construction
is based on an algebraic curve $\Sigma$ with two fixed points $P_\pm$
endowed with a function $x=x(P)$ which has simple poles at
$P=Q_1,\dots,Q_M$ where
$M$ is the complex ``dimension" of sigma-model, i.e. the number of real
co-ordinates in (\ref{sigmamo}) is $2M$.
For the $SU(2)$ case and sphere $S^3$ one should take $M=2$.

These generic solutions are expressed in terms of the
Riemann theta-functions of $\Gamma$, which is the double-cover
of $\Sigma$ with the branch points $P_\pm$, or in terms of
the Prym theta-functions on $\Sigma$ itself. The basic idea is that
the sigma-model co-ordinates are determined by the values of the
Baker-Akhiezer functions $\Upsilon (P,\sigma_\pm)$, satisfying
\be
\label{wave}
(\d_+\d_- + u)\Upsilon (P,\sigma_\pm)=0, \ \ \ \ P\in\Gamma
\ee
where $u$ is the Lagrange multiplyer of the sigma-model (\ref{sigmamo}),
which ensures equality (\ref{sphere}). The solution for the sigma-model
co-ordinates is given by
\be
\label{zkri}
Z_I(\sigma_\pm) = r_I\Upsilon(q_I,\sigma_\pm)
\ \ \ \ \
{\bar Z}_I(\sigma_\pm) = r_I\Upsilon({\bar q}_I,\sigma_\pm)
= r_I\overline{\Upsilon(q_I,\sigma_\pm)},
\ \ \ \ \
I=1,\dots,M
\ee
where $\sigma_\pm = \tau\pm\sigma$ are relativistic co-ordinates on
string world-sheet and
$Z_1=X_1+iX_2$, ${\bar Z}_1=X_1-iX_2$, {\em etc} are complexified
target-space co-ordinates of the
sigma-model. In (\ref{zkri}) $(q_I,{\bar q}_I)$ are pairs of points on
$\Gamma$ conjugated by the involution $f$, such that
$f:\Gamma\mapsto\Gamma$, $f^2=1$, and
\be
\label{invol}
\Gamma /f = \Sigma,
\ \ \ \ \ \
f: P_\pm \mapsto P_\pm,
\\
f: (q_I,{\bar q}_I) \mapsto ({\bar q}_I,q_I), \ \ \ \ I=1,\dots,M
\\
\pi_{\Gamma\to\Sigma} : (q_I,{\bar q}_I) \mapsto Q_I, \ \ \ \ I=1,\dots,M
\ee
i.e. $\Sigma$ is factor of $\Gamma$ under the action of $f$, the points
$P_+$ and $P_-$ are mapped by $f$ to themselves, while the points $q_I$ and
$\bar q_I$ (whose projections on $\Sigma$ are poles $\{ Q_I\}$ of the
function $x$ on $\Sigma$ are mapped by $f$ to each other.

The solution $\Upsilon(P,\sigma_\pm)$ to (\ref{wave})
is the Baker-Akhiezer function on $\Gamma$ with
essential singularities at $P_\pm$
\be
\Upsilon(P,\sigma_\pm) \propto \exp\left(\Omega_+(P)\sigma_+
+ \Omega_-(P)\sigma_-\right)
\theta \left({\bf{\cal A}}(P)+ {\bf U}_+\sigma_+ + {\bf U}_-\sigma_-\right) \sim
\\
\ \stackreb{P\to P_\pm}{\sim}\ \exp\left(k_\pm \sigma_\pm\right)
\left(1+\sum_{j=1}^\infty \xi_j^\pm(\sigma_\pm) k_\pm^{-j}\right)
\ee
where one can use the standard formulas (see, e.g. \cite{DKN}) with
the inverse local co-ordinates
$k_\pm(P_\pm)=\infty$, the Abel map of $\Gamma$ into its Jacobian
${\bf{\cal A}}(P) = \int^P d\bomega$, and
${\bf U}_\pm = \oint_{\bf B} d\Omega_\pm$ are ${\bf B}$-periods of the
second-kind Abelian differentials on $\Gamma$
\be
d\Omega_\pm\ \stackreb{P\to P_\pm}{\sim}\
\pm dk_\pm\left(1+O(k_\pm^{-2})\right),
\ \ \ \ \ \oint_{\bf A}d\Omega_\pm = 0
\ee
with the only second-order pole at
$P_\pm$ respectively.

The normalization factors in (\ref{zkri}) are determined by
\be
\label{ris}
r_I^2 = {\res_{Q_I} xd\Omega\over x_--x_+}, \ \ \ \ I=1,\dots,M
\ee
where $d\Omega$ is the third-kind Abelian differential on $\Sigma$
with the simple poles at $P_\pm$, $x_\pm = x(P_\pm)=\res_{P_\pm}xd\Omega$ and
normalizations (\ref{ris}) satisfy $\sum_I r_I^2=1$ due to $\sum \res
\left(xd\Omega\right) = 0$.

The Virasoro constraints (\ref{virasoro}) are ensured by
\be
\label{virkri}
\sum_I\left|\d_\pm Z_I\right|^2 = \sum_I \res_{Q_I}
\left({x-x_\pm\over x_--x_+}\d_\pm\Upsilon\d_\pm\bar\Upsilon d\Omega\right)
= \res_{P_\pm}
\left({x-x_\pm\over x_+-x_-}
\d_\pm\Upsilon\d_\pm\bar\Upsilon d\Omega\right) =
\\
= \pm 2\res_{k_\pm=\infty}\left({x-x_\pm\over x_+-x_-}\ k_\pm^2\
{dk_\pm\over k_\pm}\right) = \kappa^2
\ee
which is guaranteed after we require at $P_\pm$
\be
\label{eppm}
x(P)\ \stackreb{P\to P_\pm}{\sim}\ x_\pm \mp {\kappa^2\over 2}(x_+-x_-)
{1\over k_\pm^2} + O\left({1\over k_\pm^4}\right)
\ee
instead of used in \cite{Krisig} $\left.dx\right|_{P_\pm}=0$ which
would be equivalent to $\kappa=0$ in (\ref{virasoro}).
For the sigma-model on $S^3$, considered in sect.~\ref{ss:chiral} we
have $M=2$ and the role of $\Sigma$ is played by hyperelliptic
curve (\ref{sigmaxxx}),
and the function $x$ can be taken to be the hyperelliptic co-ordinate.

The double covers with only two
branch points have relatively simple structure (see sect.~5 of
\cite{Fay} for details).
The period matrix of $\Gamma$ is of the size $2g\times 2g$, where
$g_\Sigma=g$ and $g_\Gamma=2g$, and has the form
\be
\label{pegamma}
T_\Gamma = \ha\left(
\begin{array}{cc}
  B+T & B-T \\
  B-T & B+T
\end{array}\right)
\ee
where $T=T_\Sigma$ is $g\times g$ period matrix of $\Sigma$ and $B$ is
$g\times g$ Prym matrix of $\Sigma$.
In the ``non-relativistic" limit $P_\pm \rightarrow P$, the cut on $\Gamma$
between $P_+$ and $P_-$ shrinks to a single point, and $\Gamma$ turns
essentially into two copies of $\Sigma$. In this limit \cite{Fay}
$B=T+O(a)$, where $a$ is some parameter, measuring distance between $P_+$
and $P_-$, so the Prym
theta-functions turn into the ordinary Riemann theta-functions on $\Sigma$.

In the simplest case when $\Sigma$ is a rational curve there are no theta
functions and the Baker-Akhiezer functions are simply
\be
\label{psira}
\Upsilon = \exp\left(k_+\sigma_+ + k_-\sigma_-\right),
\ \ \ \
\bar\Upsilon = \exp\left(-k_+\sigma_+ - k_-\sigma_-\right),
\ee
where the local co-ordinates on sphere
\be
\label{kraW}
k_+ = {\kappa\over\sqrt{W}},
\ \ \ \
k_- = \kappa\sqrt{W},
\ee
are expressed through a uniformizing function $W$, such that
$W(P_+)=0$ and $W(P_-)=\infty$. Then the Abelian
differential of the third kind is $d\Omega={dW\over W}$
and generic form of the function $x$ with two poles is
\be
\label{Erat}
x = \sum_{I=1,2} {A_I\over W-W_I} + x_0
\ee
To get the desired form of expansion at zero and infinity (\ref{eppm}) one
should impose the constraint
\be
\sum_{I=1,2} {A_I\over W_I}\left(W_I-{1\over W_I}\right) = 0
\ee
and with these data one gets from (\ref{ris})
\be
\label{rrat}
r_+^2 = {{A_+\over W_+}\over {A_+\over W_+}+{A_-\over W_-}} =
{W_- -{1\over W_-}\over W_- -{1\over W_-} - W_+ +{1\over W_+}}
\\
r_-^2 = {{A_-\over W_-}\over {A_+\over W_+}+{A_-\over W_-}} =
{W_+ -{1\over W_+}\over W_+ -{1\over W_+} - W_- +{1\over W_-}}
\ee
The function (\ref{Erat}) can be then written as
\be
\label{Erat1}
x = {W_+\over W-W_+}\left(W_--{1\over W_-}\right) -
{W_-\over W-W_-}\left(W_+-{1\over W_+}\right) + x_0
\ee
If we parameterize sphere, as in (\ref{racu})
\footnote{Compare to sect.~\ref{ss:rationals} we use here the
coefficients $b=B/A$ and $c=C/A$, (cf. with \rf{Geq}) and we recall that
$a={1\over 4\pi\kappa}$.}
\be
\label{racu1}
y^2=x^2+bx+c
\ee
the uniformizing function $W$ can be chosen as
\be
\label{unif}
{W\over W_0} = {x-a+y-y_+\over x+a+y-y_-} \equiv {z-z_+\over z-z_-}
\ee
with $y_+=y(a)$, $y_-=y(-a)$, $z_\pm=y_\pm\pm a$ and a useful
co-ordinate $z$ is defined by
\be
z=x+y,
\ \ \ \ \
x={z^2-c\over b+2z},
\ \ \ \ \
y = {z^2+bz+c\over b+2z}
\ee
The constant $W_0$ is convenient to choose as
\be
\label{W0}
W_0^2={y_+(b+2z_-)\over y_-(b+2z_+)}
\ee
and the poles in (\ref{Erat}), (\ref{Erat1}) to put into
\be
W_+= W_0{b+2z_+\over b+2z_-},
\ \ \ \ \
W_-=W_0,
\ee
corresponding to two infinities $x=\infty$ of (\ref{racu1}).

The same function (\ref{unif}) with zero at $x=a$ and pole at $x=-a$ on the
``lower" or ``unphysical" sheet (corresponding in our notations
to $y = +\sqrt{x^2+bx+c}$) can be presented as
\be
\label{unif1}
{W\over W_0}
= {b+z_+ +z_-\over b+2z_-} - {z_+ -z_-\over
b+2z_-}\ {y+y_-\over x+a}
\ee
or
\be
\label{unif2}
{W_0\over W}
= {b+z_+ +z_-\over b+2z_+} + {z_+ -z_-\over
b+2z_+}\ {y+y_+\over x-a}
\ee
The resolvent \rf{Geq} can be presented as
\be
\label{GeqW}
G = -{1\over 4(z_+-z_-)}\sqrt{(b+2z_+)(b+2z_-)\over y_+y_-}
\left(W-{1\over W}\right) - {(y_+-y_-)(b+z_++z_-)\over 4y_+y_-(z_+-z_-)} =
\\
= {1\over 4}\left({1\over x-a}\left(1+{y\over y_+}\right)
+ {1\over x+a}\left(1+{y\over y_-}\right)\right)
- {(y_+-y_-)(b+z_++z_-)\over 4y_+y_-(z_+-z_-)}
\ee
The normalization coefficients (\ref{rrat}) are
determined by (\ref{ris}) with $x$ being ``hyperelliptic" co-ordinate
on \rf{racu1}
\be
r_I^2 = -{1\over 2a}\ \res_{\infty_I} \left(x{dW\over W}\right) =
\left\{
\begin{array}{c}
  \ha+{y_+-y_-\over 4a} = \ha+{b\over 4\sqrt{c}}+O(a), \ \ \ \ I=+ \\
  \ha-{y_+-y_-\over 4a} = \ha-{b\over 4\sqrt{c}}+O(a), \ \ \ \ I=-
\end{array}\right.
\ee
where $I=+,-$ and $\infty_I$ correspond to two points with $x=\infty$ on
(\ref{racu1}). From (\ref{unif}) one can easily find that
\be
x-a \stackreb{x\to a}{\sim} {2y_+(z_+-z_-)\over b+2z_+}
{W\over W_0} +
O(W^2) = 2(z_+-z_-)\sqrt{y_+y_-\over (b+2z_+)(b+2z_-)}
{\kappa^2\over k_+^2} + \dots
\\
x+a \stackreb{x\to -a}{\sim} -{2y_-(z_+-z_-)\over b+2z_-}
{W_0\over W} +
O(W^{-2}) = -2(z_+-z_-)\sqrt{y_+y_-\over (b+2z_+)(b+2z_-)}
{\kappa^2\over k_-^2} + \dots
\ee
i.e. the ``hyperelliptic" co-ordinate $x$ (the normalization here is
inessential) indeed satisfies (\ref{eppm}).

\sectiono{Elliptic integrals
\label{ss:ellint}}

We use the following conventions for elliptic integrals:
\be\label{ellint}
K(r) = \int_{0}^{1} \frac{dx}{\sqrt{(1-x^2)(1-rx^2)}}\,,
\nonumber  \\*
E(r) = \int_{0}^{1} dx\,\sqrt{\frac{1-rx^2}{1-x^2}}\,.
\ee
Here we list some of their properties \cite{BE} which are used
in the main text.

Under the modular transformation $r\rightarrow 1/r$
elliptic integrals transform as follows
\be \label{Landen}
K(1/r)=\sqrt{r}\left(K(r)-iK(1-r)\right),
\nonumber \\
K(1-1/r)=\sqrt{r}K(1-r),
\nonumber \\
E(1/r)=\frac{E(r)+iE(1-r)-(1-r)K(r)-irK(1-r)}{\sqrt{r}}\,,
\nonumber \\
E(1-1/r)=\frac{E(1-r)}{\sqrt{r}}\,.
\ee
The elliptic integrals of direct and dual moduli
satisfy the Legendre relation:
\begin{equation}\label{Legendre}
K(r)E(1-r)-K(r)K(1-r)+E(r)K(1-r)=\frac{\pi}{2}\,.
\end{equation}

\sectiono{Reality condition for elliptic solutions
\label{ss:real}}

The end-points of the cuts for the elliptic solution of Bethe equations
must be complex conjugate of one another. This is not
always the case for solutions
of (\ref{1-2alpha}), (\ref{ab}). The reality condition thus puts a
constraint on eligible $m$ and $n$. We know that the solutions are real
for $m=0$ or $n=0$ (cases (i) and (ii) in sect.~\ref{ss:ellip}),
since they reduce to the solutions found in  \cite{Beisert:2003xu}.
 Here we analyze solutions with arbitrary
$n$ and $m$ and assume that $m\neq 0$. We will analyze
the reality condition in the limit of
$\alpha\rightarrow 0$.
Then $r$ is exponentially
close to one and we can expand elliptic integrals as follows:
\be
K(r)=\varepsilon+4(\varepsilon-1)\e^{-2\varepsilon}+O\left(\varepsilon^2\e^{-4\varepsilon}\right),
\nonumber \\
K(1-r)=\frac{\pi}{2}+2\pi\e^{-2\varepsilon}+O\left(\e^{-4\varepsilon}\right),
\nonumber \\
E(r)=1+4(2\varepsilon-1)\e^{-2\varepsilon}+O\left(\varepsilon^2\e^{-4\varepsilon}\right),
\nonumber \\
E(1-r)=\frac{\pi}{2}-2\pi\e^{-2\varepsilon}+O\left(\e^{-4\varepsilon}\right),
\ee
where $r=1-16\e^{-2\varepsilon}$.
Plugging these expansions into (\ref{1-2alpha}), we find
\be
\varepsilon+\frac{i\pi n}{2m}=\frac{1}{2\alpha}\left[1+4\e^{\frac{i\pi
n}{m}-\frac{1}{\alpha}}
+O\left(\e^{-\frac{2}{\alpha}}\right)\right],
\ee
and, from (\ref{ab}):
\be
{\rm x}_1=\frac{i\alpha}{2m}\left[1-4\e^{\frac{i\pi n}{m}-\frac{1}{\alpha}}
+O\left(\e^{-\frac{2}{\alpha}}\right)\right],
\nonumber \\
{\rm x}_2=\frac{i\alpha}{2m}\left[1+4\e^{\frac{i\pi n}{m}-\frac{1}{\alpha}}
+O\left(\e^{-\frac{2}{\alpha}}\right)\right].
\ee
There are only two ways to satisfy the reality condition.
We can either put $n=0$, then ${\rm x}_1=-\overline{{\rm x}_1}$
and ${\rm x}_2=-\overline{{\rm x}_2}$, which corresponds to case (i) in sect.~\ref{ss:ellip},
or we can  take $m=2n$, then ${\rm x}_1=-\overline{{\rm x}_2}$, which corresponds to case (iii).


\begin{thebibliography}{99}
\addtolength{\itemsep}{-6pt}

\bibitem{Polyakov}
A.~M.~Polyakov, {\it Gauge Fields and Strings},
Harwood Academic Publishers, 1987.

\bibitem{'tHooft:1973jz}
G.~'t Hooft,
``A Planar Diagram Theory For Strong Interactions,''
Nucl.\ Phys.\ B {\bf 72}, 461 (1974).



\bibitem{David:nj}
F.~David,
``A Model Of Random Surfaces With Nontrivial Critical Behavior,''
Nucl.\ Phys.\ B {\bf 257}, 543 (1985).


\bibitem{Kazakov:ds}
V.~A.~Kazakov,
``Bilocal Regularization Of Models Of Random Surfaces,''
Phys.\ Lett.\ B {\bf 150}, 282 (1985).


\bibitem{Brezin:1977sv}
E.~Brezin, C.~Itzykson, G.~Parisi and J.~B.~Zuber,
``Planar Diagrams,''
Commun.\ Math.\ Phys.\  {\bf 59}, 35 (1978).

\bibitem{'tHooft:1974hx}
G.~'t Hooft,
``A Two-Dimensional Model For Mesons,''
Nucl.\ Phys.\ B {\bf 75}, 461 (1974).


\bibitem{Maldacena:1998re}
J.~M.~Maldacena,
{``The large N limit of superconformal field theories and
  supergravity''},
Adv.~Theor.~Math.~Phys. ~\textbf{2}~(1998)~231,
[arXiv:hep-th/9711200].

\bibitem{Gubser:1998bc}
S.~S.~Gubser, I.~R.~Klebanov and A.~M.~Polyakov,
{``Gauge theory correlators from non-critical string theory''},
Phys.~Lett.~\textbf{B428}~(1998)~105,
[arXiv:hep-th/9802109].

\bibitem{Witten:1998qj}
E.~Witten,
{``Anti-de Sitter space and holography''},
Adv.~Theor.~Math.~Phys.~\textbf{2}~(1998)~253,
[arXiv:hep-th/9802150].

\bibitem{Berenstein:2002jq}
D.~Berenstein, J.~M.~Maldacena and H.~Nastase,
{``Strings in flat space and pp waves from {$\mathcal{N}=\mathord{}$4}
  {Super} {Yang Mills}''},
JHEP~\textbf{0204}~(2002)~013,
[arXiv:hep-th/0202021].


\bibitem{Gubser:2002tv}
S.~S.~Gubser, I.~R.~Klebanov and A.~M.~Polyakov,
{``A semi-classical limit of the gauge/string correspondence''},
Nucl.~Phys.~\textbf{B636}~(2002)~99,
[arXiv:hep-th/0204051].

\bibitem{Metsaev:2001bj}
R.~R.~Metsaev,
{``Type {IIB} {Green-Schwarz} superstring in plane wave {Ramond-Ramond}
  background''},
Nucl.~Phys.~\textbf{B625}~(2002)~70,
[arXiv:hep-th/0112044].

\bibitem{Metsaev:2002re}
R.~R.~Metsaev and A.~A.~Tseytlin,
{``Exactly solvable model of superstring in plane wave {Ramond-Ramond}
  background''},
Phys.~Rev.~\textbf{D65}~(2002)~126004,
[arXiv:hep-th/0202109].

\bibitem{Gross:2002su}
D.~J.~Gross, A.~Mikhailov and R.~Roiban,
{``Operators with large R charge in {$\mathcal{N}=\mathord{}$4}
  Yang-Mills theory''},
Annals~Phys.~\textbf{301}~(2002)~31,
[arXiv:hep-th/0205066].

\bibitem{Santambrogio:2002sb}
A.~Santambrogio and D.~Zanon,
{``Exact anomalous dimensions of \4N Yang-Mills
  operators with large R charge''},
Phys.~Lett.~\textbf{B545}~(2002)~425,
[arXiv:hep-th/0206079].

\bibitem{Frolov:2002av}
S.~Frolov and A.~A.~Tseytlin,
{``Semiclassical quantization of rotating superstring in {$AdS_5 \times
  S^5$}''},
JHEP~\textbf{0206}~(2002)~007,
[arXiv:hep-th/0204226].

\bibitem{Russo:2002sr}
J.~G.~Russo,
{``Anomalous dimensions in gauge theories from rotating strings in
  {$AdS_5 \times S^5$}''},
JHEP~\textbf{0206}~(2002)~038,
[arXiv:hep-th/0205244].

\bibitem{Minahan:2002rc}
J.~A.~Minahan,
{``Circular semiclassical string solutions on {$AdS_5\times S^5$}''},
Nucl.~Phys.~\textbf{B648}~(2003)~203,
[arXiv:hep-th/0209047].
%
\bibitem{Tseytlin:2002ny}
A.~A.~Tseytlin,
{``Semiclassical quantization of superstrings: {$AdS_5\times S^5$} and
  beyond''},
Int.~J.~Mod.~Phys. \textbf{A18}~(2003)~981,
[arXiv:hep-th/0209116].

\bibitem{Frolov:2003qc}
S.~Frolov and A.~A.~Tseytlin,
{``Multi-spin string solutions in {$AdS_5\times S^5$}''},
Nucl.~Phys.~\textbf{B668}~(2003)~77,
[arXiv:hep-th/0304255].
%
\bibitem{Frolov:2003tu}
S.~Frolov and A.~A.~Tseytlin,
{``Quantizing three-spin string solution in {$AdS_5 \times S^5$}''},
JHEP~\textbf{0307}~(2003)~016,
[arXiv:hep-th/0306130].
%

\bibitem{Frolov:2003xy}
S.~Frolov and A.~A.~Tseytlin,
``Rotating string solutions: AdS/CFT duality in non-supersymmetric  sectors,''
Phys.\ Lett.\ B {\bf 570}, 96 (2003)
[arXiv:hep-th/0306143].

\bibitem{Arutyunov:2003uj}
G.~Arutyunov, S.~Frolov, J.~Russo and A.~A.~Tseytlin,
{``Spinning strings in $AdS_5\times S^5$ and integrable systems''},
[arXiv:hep-th/0307191].
%
\bibitem{Engquist:2003rn}
J.~Engquist, J.~A.~Minahan and K.~Zarembo,
``Yang-Mills duals for semiclassical strings on $AdS_5\times S^5$,''
JHEP {\bf 0311} (2003) 063
[arXiv:hep-th/0310188].

\bibitem{Arutyunov:2003za}
G.~Arutyunov, J.~Russo and A.~A.~Tseytlin,
``Spinning strings in $AdS_5\times S^5$: New integrable system relations,''
arXiv:hep-th/0311004.

\bibitem{Tseytlin:2003ii}
A.~A.~Tseytlin,
``Spinning strings and AdS/CFT duality,''
arXiv:hep-th/0311139.


\bibitem{Minahan:2002ve}
J.~A.~Minahan and K.~Zarembo,
{``The Bethe-ansatz for \4N super Yang-Mills,''}
JHEP {\bf 0303}, 013 (2003)
[arXiv:hep-th/0212208].

\bibitem{Beisert:2003jj}
N.~Beisert,
{``The complete one-loop dilatation operator of \4N super Yang-Mills
theory,''}
arXiv:hep-th/0307015.

\bibitem{Beisert:2003yb}
N.~Beisert and M.~Staudacher,
{``The \4N SYM integrable super spin chain,''}
Nucl.\ Phys.\ B {\bf 670}, 439 (2003)
[arXiv:hep-th/0307042].

\bibitem{Bethe:1931hc}
H.~Bethe,
``On The Theory Of Metals. 1. Eigenvalues And Eigenfunctions For The Linear
Atomic Chain,''
Z.\ Phys.\  {\bf 71}, 205 (1931).

\bibitem{Faddeev:1996iy}
L.~D.~Faddeev,
``How Algebraic Bethe Ansatz works for integrable model,''
arXiv:hep-th/9605187.

\bibitem{Beisert:2003xu}
N.~Beisert, J.~A.~Minahan, M.~Staudacher and K.~Zarembo,
``Stringing spins and spinning strings,''
JHEP {\bf 0309} (2003) 010
[arXiv:hep-th/0306139].

\bibitem{Beisert:2003ea}
N.~Beisert, S.~Frolov, M.~Staudacher and A.~A.~Tseytlin,
``Precision spectroscopy of AdS/CFT,''
JHEP {\bf 0310}, 037 (2003)
[arXiv:hep-th/0308117].

\bibitem{Kristjansen:2004ei}
C.~Kristjansen,
``Three-spin Strings on $AdS_5\times S^5$ from \4N SYM,''
arXiv:hep-th/0402033.

\bibitem{Kruczenski:2003gt}
M.~Kruczenski,
``Spin chains and string theory,''
arXiv:hep-th/0311203.

\bibitem{RESHSM}
N. Reshetikhin and F. Smirnov, ``Quantum Floquet functions'',
Zapiski nauchnikh seminarov LOMI (Notes of scientific seminars of
Leningrad Branch of Steklov Institute) v.131 (1983) 128 (in russian).


\bibitem{Smirnov:1998kv}
F.~A.~Smirnov,
``Quasi-classical study of form factors in finite volume,''
arXiv:hep-th/9802132.


\bibitem{Beisert:2003tq}
N.~Beisert, C.~Kristjansen and M.~Staudacher,
``The dilatation operator of \4N super Yang-Mills theory,''
Nucl.\ Phys.\ B {\bf 664}, 131 (2003)
[arXiv:hep-th/0303060].

\bibitem{Beisert:2003jb}
N.~Beisert,
``Higher loops, integrability and the near BMN limit,''
JHEP {\bf 0309}, 062 (2003)
[arXiv:hep-th/0308074].

\bibitem{Beisert:2003ys}
N.~Beisert,
``The su(2$|$3) dynamic spin chain,''
arXiv:hep-th/0310252.

\bibitem{Klose:2003qc}
T.~Klose and J.~Plefka,
``On the integrability of large N plane-wave matrix theory,''
Nucl.\ Phys.\ B {\bf 679}, 127 (2004)
[arXiv:hep-th/0310232].

\bibitem{Serban:2004jf}
D.~Serban and M.~Staudacher,
{``Planar \4N gauge theory and the Inozemtsev long range spin
chain''},
arXiv:hep-th/0401057.

\bibitem{Bena:2003wd}
I.~Bena, J.~Polchinski and R.~Roiban,
 ``Hidden symmetries of the $AdS_5\times S^5$ superstring,''
arXiv:hep-th/0305116.

\bibitem{Mandal:2002fs}
G.~Mandal, N.~V.~Suryanarayana and S.~R.~Wadia,
 ``Aspects of semiclassical strings in $AdS_5$,''
Phys.\ Lett.\ B {\bf 543}, 81 (2002)
[arXiv:hep-th/0206103].

\bibitem{Dolan:2003uh}
L.~Dolan, C.~R.~Nappi and E.~Witten,
``A relation between approaches to integrability in superconformal Yang-Mills
theory,''
JHEP {\bf 0310}, 017 (2003)
[arXiv:hep-th/0308089];
``Yangian Symmetry in D=4 Superconformal Yang-Mills Theory,''
arXiv:hep-th/0401243.

\bibitem{Arutyunov:2003rg}
G.~Arutyunov and M.~Staudacher,
``Matching higher conserved charges for strings and spins,''
arXiv:hep-th/0310182.

\bibitem{Engquist:2004bx}
J.~Engquist,
 ``Higher Conserved Charges and Integrability for Spinning Strings in
$AdS_5 \times S^5$,''
arXiv:hep-th/0402092.

\bibitem{Inozemtsev}
V.~I.~Inozemtsev, ``On the connection between the one-dimensional
s=1/2 Heisenberg chain and Haldane-Shastry model'', J.Stat.Phys {\bf
59} (1990) 1143;
 ``Integrable Heisenberg-van Vleck chains with
variable range exchange,'' Phys.\ Part.\ Nucl.\ {\bf 34}, 166 (2003)
[Fiz.\ Elem.\ Chast.\ Atom.\ Yadra {\bf 34}, 332 (2003)]
[arXiv:hep-th/0201001].

\bibitem{SW}
N.~Seiberg and E.~Witten, Nucl. Phys. {\bf B426} (1994) 19,
hep-th/ 9407087.

\bibitem{GKMMM}
A.~Gorsky, I.~Krichever, A.~Marshakov, A.~Mironov and A.~Morozov, Phys. Lett.
{\bf B355} (1995) 466; hep-th/9505035.

\bibitem{Lipatov:1993yb}
L.~N.~Lipatov,
``High-energy asymptotics of multicolor QCD and exactly solvable lattice
models,''
JETP Lett.\  {\bf 59}, 596 (1994)
[Pisma Zh.\ Eksp.\ Teor.\ Fiz.\  {\bf 59}, 571 (1994)]
[arXiv:hep-th/9311037].

\bibitem{Faddeev:1994zg}
L.~D.~Faddeev and G.~P.~Korchemsky,
``High-energy QCD as a completely integrable model,''
Phys.\ Lett.\ B {\bf 342}, 311 (1995)
[arXiv:hep-th/9404173].

\bibitem{Braun:1998id}
V.~M.~Braun, S.~E.~Derkachov and A.~N.~Manashov,
``Integrability of three-particle evolution equations in {QCD},''
Phys.\ Rev.\ Lett.\  {\bf 81}, 2020 (1998)
[arXiv:hep-ph/9805225].

\bibitem{Braun:1999te}
V.~M.~Braun, S.~E.~Derkachov, G.~P.~Korchemsky and A.~N.~Manashov,
``Baryon distribution amplitudes in {QCD},''

Nucl.\ Phys.\ B {\bf 553}, 355 (1999)
[arXiv:hep-ph/9902375].

\bibitem{Belitsky:2003ys}
A.~V.~Belitsky, A.~S.~Gorsky and G.~P.~Korchemsky,
``Gauge / string duality for QCD conformal operators,''
Nucl.\ Phys.\ B {\bf 667}, 3 (2003)
[arXiv:hep-th/0304028].

\bibitem{KorchQ}
G.~P.~Korchemsky,
``Quasiclassical QCD pomeron'',
Nucl.\ Phys.\ B {\bf 462}, 333 (1996)
[arXiv:hep-th/9508025];
``Integrable structures and duality in high-energy QCD",
Nucl.\ Phys.\ B {\bf 498} (1997) 68
[arXiv:hep-th/9609123].

\bibitem{KorchKri}
G.~P.~Korchemsky, I.~M.~Krichever,
"Solitons in high-energy QCD",
Nucl.\ Phys.\ B {\bf 505} 387 (1997),
[arXiv:hep-th/9704079].


\bibitem{Wang:2003cu}
X.~J.~Wang and Y.~S.~Wu,
 ``Integrable spin chain and operator mixing in N = 1,2 supersymmetric
arXiv:hep-th/0311073.

\bibitem{Stefanski:2003qr}
B.~.~J.~Stefanski,
``Open spinning strings,''
arXiv:hep-th/0312091.

\bibitem{Roiban:2003dw}
R.~Roiban,
``On spin chains and field theories,''
arXiv:hep-th/0312218.

\bibitem{Chen:2004mu}
B.~Chen, X.~J.~Wang and Y.~S.~Wu,
 ``Integrable open spin chain in super Yang-Mills and the plane-wave/SYM
duality,''
arXiv:hep-th/0401016.

\bibitem{DeWolfe:2004zt}
O.~DeWolfe and N.~Mann,
``Integrable open spin chains in defect conformal field theory,''
arXiv:hep-th/0401041.

\bibitem{Gorsky:2003nq}
A.~Gorsky,
``Spin chains and gauge / string duality,''
arXiv:hep-th/0308182.

\bibitem{Douglas:1993ii}
M.~R.~Douglas and V.~A.~Kazakov,
Phys.\ Lett.\ B {\bf 319}, 219 (1993)
[arXiv:hep-th/9305047].


\bibitem{MATTHIAS}
M.~Staudacher, private communication;
J.~Minahan, unpublished.

\bibitem{David:sk}
F.~David,
``Phases Of The Large N Matrix Model And Nonperturbative Effects In 2-D
Gravity,''
Nucl.\ Phys.\ B {\bf 348}, 507 (1991).

\bibitem{KM}
V.~A.~Kazakov and A.~Marshakov,
``Complex curve of the two matrix model and its tau-function,''
J.\ Phys.\ A {\bf 36}, 3107 (2003)
[arXiv:hep-th/0211236].

\bibitem{Felder:2004uy}
G.~Felder and R.~Riser,
``Holomorphic matrix integrals,''
arXiv:hep-th/0401191.

\bibitem{Pohlmeyer:1975nb}
K.~Pohlmeyer,
``Integrable Hamiltonian Systems And Interactions Through Quadratic
Constraints,''
Commun.\ Math.\ Phys.\  {\bf 46}, 207 (1976).

\bibitem{Zakharov:pp}
V.~E.~Zakharov and A.~V.~Mikhailov,
``Relativistically Invariant Two-Dimensional Models In Field Theory Integrable
By The Inverse Problem Technique. (In Russian),''
Sov.\ Phys.\ JETP {\bf 47}, 1017 (1978)
[Zh.\ Eksp.\ Teor.\ Fiz.\  {\bf 74}, 1953 (1978)].



\bibitem{book_of_soliton}
 S.P. Novikov, S.V. Manakov, L.P. Pitaevskii and V.E. Zakharov,
{\it Theory of solitons : the inverse scattering method,}
(Consultants Bureau, 1984).

\bibitem{DKN}
B.Dubrovin, I.Krichever and S.Novikov, {\it Integrable systems - I},
{\sl Sovremennye problemy matematiki (VINITI), Dynamical systems - 4}
(1985) 179 (in Russian).

\bibitem{Faddeev's_book}
 L.D. Faddeev and L.A. Takhtajan,
{\it Hamiltonian methods in the theory of solitons}
(Springer-Verlag, 1987).

\bibitem{Mikhailov:2003gq}
A.~Mikhailov,
``Speeding strings,''
JHEP {\bf 0312}, 058 (2003)
[arXiv:hep-th/0311019];
``Slow evolution of nearly-degenerate extremal surfaces,''
arXiv:hep-th/0402067.

\bibitem{Lakshmanan}
M. Lakshmanan,
``Continuum spin system as an exactly solvable dynamical system,''
Phys.\ Lett.\ A {\bf 61} (1977) 53.

\bibitem{Takhtajan:rv}
L.~A.~Takhtajan,
``Integration Of The Continuous Heisenberg Spin Chain Through The Inverse
Scattering Method,''
Phys.\ Lett.\ A {\bf 64} (1977) 235.


\bibitem{Zakharov:jc}
V.~E.~Zakharov and L.~A.~Takhtajan,
``Equivalence Of The Nonlinear Schrodinger Equation And The Equation Of A
Heisenberg Ferromagnet,''
Theor.\ Math.\ Phys.\  {\bf 38} (1979) 17
[Teor.\ Mat.\ Fiz.\  {\bf 38} (1979) 26].


\bibitem{Bob}
R.F.~Bikbaev, A.I.~Bobenko and A.R.~Its,
``Finite-zone integration of the Landau-Lifshits equation,''
Dokl. Akad. Nauk SSSR 272 (1983) 1293.


\bibitem{Faddeev:1985qu}
L.~D.~Faddeev and N.~Y.~Reshetikhin,
``Integrability Of The Principal Chiral Field Model In (1+1)-Dimension,''
Annals Phys.\  {\bf 167} (1986) 227.

\bibitem{Destri:1987hc}
C.~Destri and H.~J.~de Vega,
``Integrable Quantum Field Theories And Conformal Field Theories From Lattice
Models In The Light Cone Approach,''
Phys.\ Lett.\ B {\bf 201}, 261 (1988).

\bibitem{Callan:2003xr}
C.~G.~.~Callan, H.~K.~Lee, T.~McLoughlin, J.~H.~Schwarz, I.~Swanson and X.~Wu,
``Quantizing string theory in $AdS_5\times S^5$: Beyond the pp-wave,''
Nucl.\ Phys.\ B {\bf 673}, 3 (2003)
[arXiv:hep-th/0307032].

\bibitem{GS}
G.~Arutyunov and M.~Staudacher,
``Two-loop commuting charges and the string/gauge duality,''
arXiv:hep-th/0403077.

\bibitem{Kruczenski:2004kw}
M.~Kruczenski, A.~V.~Ryzhov and A.~A.~Tseytlin,
 ``Large spin limit of $AdS_5 \times S^5$ string theory and low energy expansion of
ferromagnetic spin chains,''
arXiv:hep-th/0403120.

\bibitem{Krisig}
I.Krichever, ``Two-dimensional algebraic-geometrical operators with
self-consistent potentials'', Func.~An.~\&~Apps. {\bf 28} (1994) No 1,
26.

\bibitem{Fay}
J.D.Fay, ``Theta Functions on Riemann Surfaces",
Lect.~Notes in Mathematics {\bf 352}, Springer-Verlag, 1973.

\bibitem{BE}
H. Bateman, A. Erdelyi,
{\it Higher Transcendental Functions} (McGraw-Hill, 1955).



\end{thebibliography}
\end{document}